# Cryostat Design

*V. Parma*[1]
CERN, Geneva, Switzerland

**Abstract**
This paper aims to give non-expert engineers and scientists working in the domain of accelerators a general introduction to the main disciplines and technologies involved in the design and construction of accelerator cryostats. Far from being an exhaustive coverage of these topics, an attempt is made to provide simple design and calculation rules for a preliminary design of cryostats. Recurrent reference is made to the Large Hadron Collider magnet cryostats, as most of the material presented is taken from their design and construction at CERN.

## 1    Introduction

The term *cryostat* (from the Greek word κρύο meaning cold, and *stat* meaning static or stable) is generally employed to describe any container housing devices or fluids kept at *very low temperatures*; the notion of 'very low temperature' generally refers to temperatures that are well below those encountered naturally on Earth, typically below 120 K.

The very first cryostats were used in the pioneering years of cryogenics as containers for liquefied gases. The *invention* of the first performing cryostats is generally attributed to Sir James Dewar (Fig. 1), and hence cryostats containing cryogenic fluids are nowadays also called *dewars*. In 1897 Dewar used silver-plated double-walled glass containers to collect the first liquefied hydrogen.

Even though the heat transfer phenomena were not well mastered during his époque, Dewar understood the benefits of thermal insulation by vacuum pumping the double-walled envelope, as well as shielding thermal radiation by silver-plating the glass walls.

H. Kamerlingh Onnes further developed *glass-blowing* which became the *enabling technology* for making dewars for his laboratory in Leiden. He introduced this technique as one of the specialities in the school of instrumentation he founded, the Leidse Instrumentmakersschool, which still exists today.

Since those times, the evolution of cryostats has been led by specific needs for the variety of applications. Today, cryostats can be found in a large range of applications spanning from industrial products to specific devices for scientific research instruments.

Two representative examples of cryostats produced in large industrial series are cryostats for superconducting magnet medical devices such as magnetic resonance imaging (MRI), with a typical worldwide production of a thousand units per year, and storage reservoirs for industrial cryogenic fluids ($LN_2$, LO, LH) produced in several thousands of units per year. Figure 2 illustrates a few of these examples.

---

[1] vittorio.parma@cern.ch

A possible breakdown of cryostats based on their applications is given in Fig. 3, where those for accelerator applications are detailed. Cryostats are needed in most particles accelerators, from linacs to synchrotron machines, where superconductivity is the enabling technology. Under test cryostats we list those that are specifically developed for testing accelerator devices, though standard commercial test cryostats are also available for a number of laboratory applications. In this course we will concentrate on cryostats for accelerator devices, housing either superconducting magnets or superconducting Radio Frequency (RF) cavities, and cooled with liquid helium.

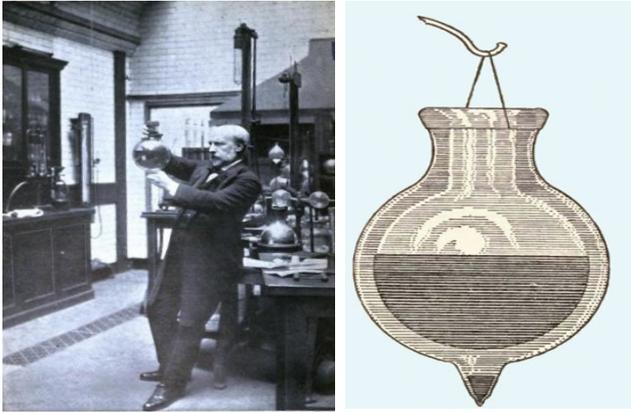

**Fig. 1:** Sir James Dewar (1842–1923), and his double-walled glass dewar

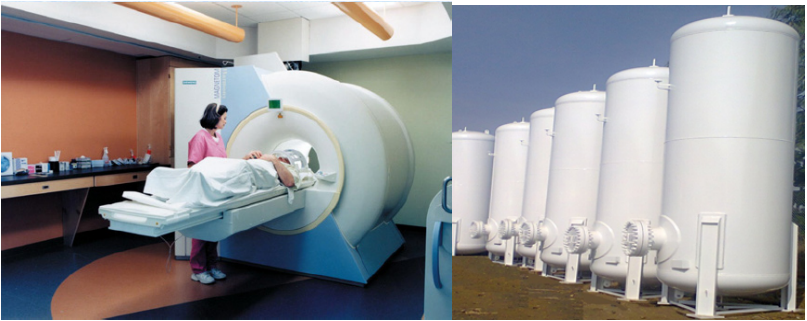

(a)                                                                    (b)

**Fig. 2:** (a) An MRI device; (b) cryogenic storage tanks

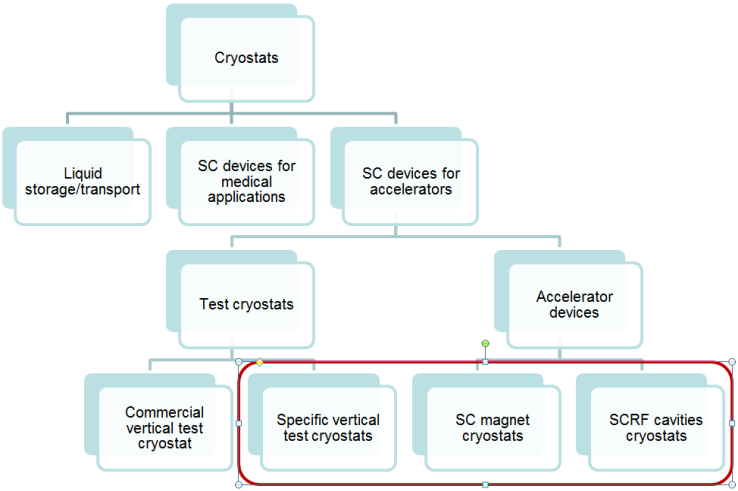

**Fig. 3:** Cryostats and their applications

## 2 Cryostat requirements

When designing cryostats for accelerator superconducting devices, the engineer is confronted with the interaction between various technical disciplines, some of which are the domain of specialists, like superconductivity, cryogenics and vacuum (Fig. 4). In these domains has to be developed a general understanding in order to be able to interact with specialists.

The basic function of a cryostat is to house and thermally insulate a superconducting device, while providing all the interfaces for its reliable and safe operation (cryogen feeding, powering, diagnostics instrumentation, safety devices, etc.) The basic technical competencies for a cryostat design engineer are mechanical engineering and heat transfer. But their application at cryogenic temperatures calls for specific competencies on thermal and mechanical properties of materials at low temperatures, which make the work very specific and a discipline in itself. Cryostat design also includes system integration of complex devices that have to be housed inside the insulation vacuum and often at cryogenic temperatures: typical examples are beam instrumentation such as Beam Position Monitors (BPM), cryogenic instrumentation (temperature, pressure, level, and mass flow sensors) and control devices (valves, servo-motors, and mechanical transmissions), electrical circuits for magnets' and cavities' powering and protection, etc. All these items or devices need to be designed for integration so that they can be assembled, operated and maintained when needed and, in most cases, designed not to hinder the overall thermal performance of the cryostats.

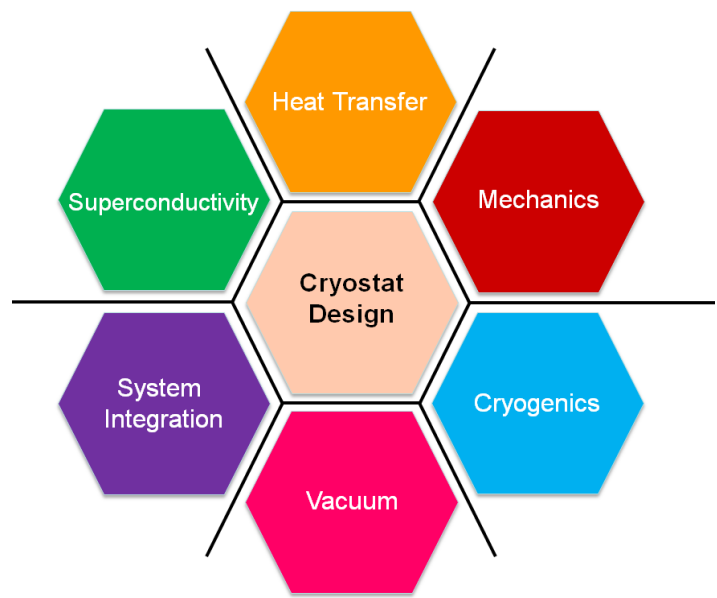

**Fig. 4:** Cryostat design, and related disciplines

An illustrative example of a cryostat for accelerators is one of the LHC dipoles (Fig. 5).

The main functionality of the cryostat is to support and position precisely the heavy dipole magnets while limiting to a minimum the thermal heat loads from ambient temperature to the 2 K temperature of the magnets. The magnets are supported on support posts made of low thermal conductivity material and are placed inside vacuum vessels in order to remove heat loads from gas convection and conduction; thermal shields cooled to a higher temperature than that of the magnets, and wrapped with reflective Multilayer Insulation (MLI), further contribute to reducing the radiation heat loads to the magnets.

The precise alignment of the magnets within the accelerator is obtained by the fine adjustment of the external jack supporting the cryo-magnet assembly while its position is measured by surveying alignment targets mounted on the vacuum vessel. The positioning accuracy and stability of the magnet inside the cryostats is the main requirement for which the supporting system has been designed.

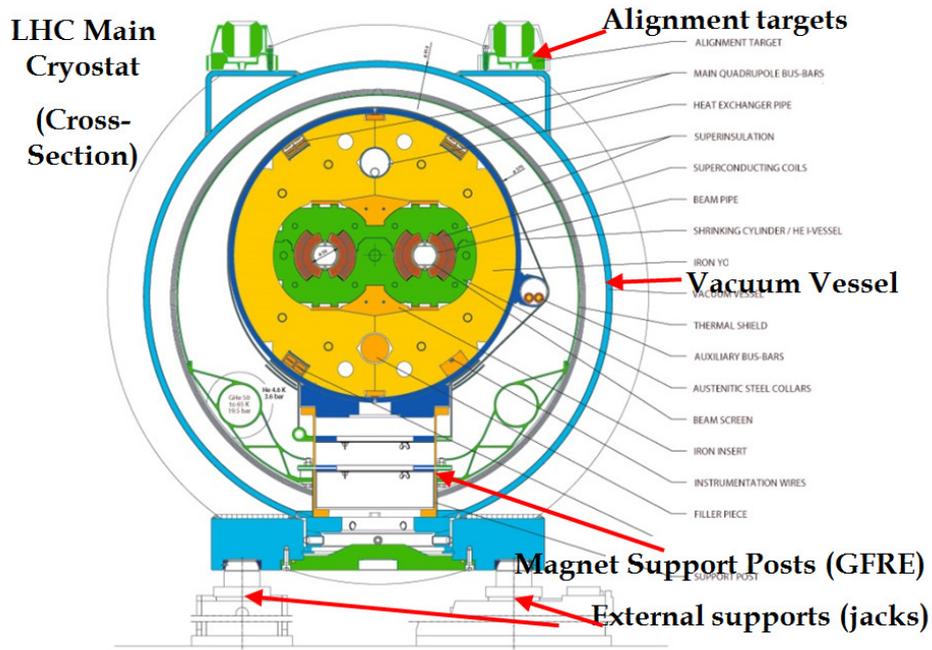

**Fig. 5:** Cross-section view of the LHC dipole cryostat

Table 1 summarises the main requirements of the supporting and alignment system for the main dipoles and quadrupoles of the LHC, where sub-millimetre positional reproducibility and stability as well as alignment precision is sought.

**Table 1:** Main requirements of the supporting and alignments system of the LHC main magnets

| Property/requirement | Dipole | Quadrupole |
|---|---|---|
| Weight | 300 kN | 65 kN |
| Cold mass in cryostat positioning accuracy (after assembly): | | |
|     Radial | ± 1 mm | ± 0.5 mm |
|     Vertical | ± 1 mm | ± 0.5 mm |
|     Longitudinal | ± 2 mm | ± 1 mm |
| Positioning reproducibility–stability of cold mass in cryostat (in operation, during lifetime): | | |
|     Radial | ± 0.3 mm (rms) | < ± 0.3 mm (rms) |
|     Vertical | ± 0.3 mm (rms) | < ± 0.3 mm (rms) |
|     Longitudinal | ± 1 mm (rms) | < ± 1 mm (rms) |
|     Radial tilt | ± 0.3 mrad (rms) | < ± 0.3 mrad (rms) |
| External supporting system: | | |
|     Alignment range | ± 20 mm | ± 20 mm |
|     Movement resolution | ± 0.05 mm | ± 0.05 mm |

Considering the extremely high electrical power consumption of the cryogenic plants for heat extraction at cryogenic temperatures (about 1 kW W$^{-1}$ from 1.8 K to 300 K), the static heat inleak to the cold masses at 1.8 K had to be minimized in the most efficient and economical way. During the course of the development of the LHC cryostats and its components, systematic laboratory heat load measurements were performed to validate designs, manufacturing and assembly methods, completed by prototype thermal performance assessments. The thermal heat loads per unit length reached for the LHC cryostats are illustrated in Table 2.

**Table 2:** Thermal budget per unit length of the LHC cryostats (W m$^{-1}$)

| | $T$ | | |
| --- | --- | --- | --- |
| | 50–75 K | 4.6–20 K | 1.9 K |
| Static heat loads (W m$^{-1}$) | 7.7 | 0.23 | 0.21 |

The following chapters present a general, but non-exhaustive, coverage of these disciplines. The LHC cryostat will often be used as a case study throughout the course. The reader is referred to the bibliography and references for further developments, see Refs. [1–18].

## 3 Heat transfer for cryostats

In the following we shall consider the heat transfer phenomena that are relevant to cryostats. *Solid conduction*, which is generally the dominating contribution in supporting systems and all cryostat feed-throughs (current leads, instrumentation wires, beam pipes, etc.); *residual gas conduction*, which can introduce a non-negligible contribution in the case of a non-perfect vacuum insulation; *thermal radiation*, which is normally the dominating heat load contribution; and the use of *multilayer insulation* and *thermal shielding* as the protection measures normally employed.

### 3.1 Solid conduction

Fourier's law expresses heat conduction in solids as being proportional to a temperature gradient through $k$, which is the *thermal conductivity* of a material:

$$\dot{Q} = -k(T) \cdot A \cdot \text{grad}(T) \, . \tag{1}$$

At cryogenic temperatures $k$ is generally temperature-dependent and non-linear, implying that numerical integration over temperature is necessary in most practical cases.

Figure 6 shows that thermal conductivity for solid materials spreads over more than five orders of magnitude, underlining the importance of the selection of materials where thermal conductivity is a property of interest. We can also notice that, for a single material, the trend is a reduction of thermal conductivity with temperature, as would be expected by the reduction with temperature of phonon vibrations in the material structure. This is, however, not the case in metals, where the main contribution to conductivity is the free movement of conduction electrons inside the crystalline lattice. In this case electron scatter with phonons reduces with temperature, hence increasing the mobility of electrons, resulting in an increase of thermal conductivity. At lower temperatures a second scatter mechanism with impurities prevails and thermal conductivity reduces once again.

In high-purity metals, the increase of thermal conductivity is very pronounced (more than one order of magnitude for pure coppers, as can be seen in Fig. 6); but in metal alloys, where impurities dominate, thermal conductivity shows a monotonic decrease according to the relation:

$$k(T) \cong T/a_i \tag{2}$$

where $a_i$ is a constant depending on the level of impurities of the material.

It is interesting to note, from electron conduction theory and in an analogy with electrical diffusion that, for metals, thermal conductivity $k$ and electrical resistivity $\sigma$ follow the Wiedemann–Franz law:

$$\frac{k}{\sigma} \cong L \cdot T \tag{3}$$

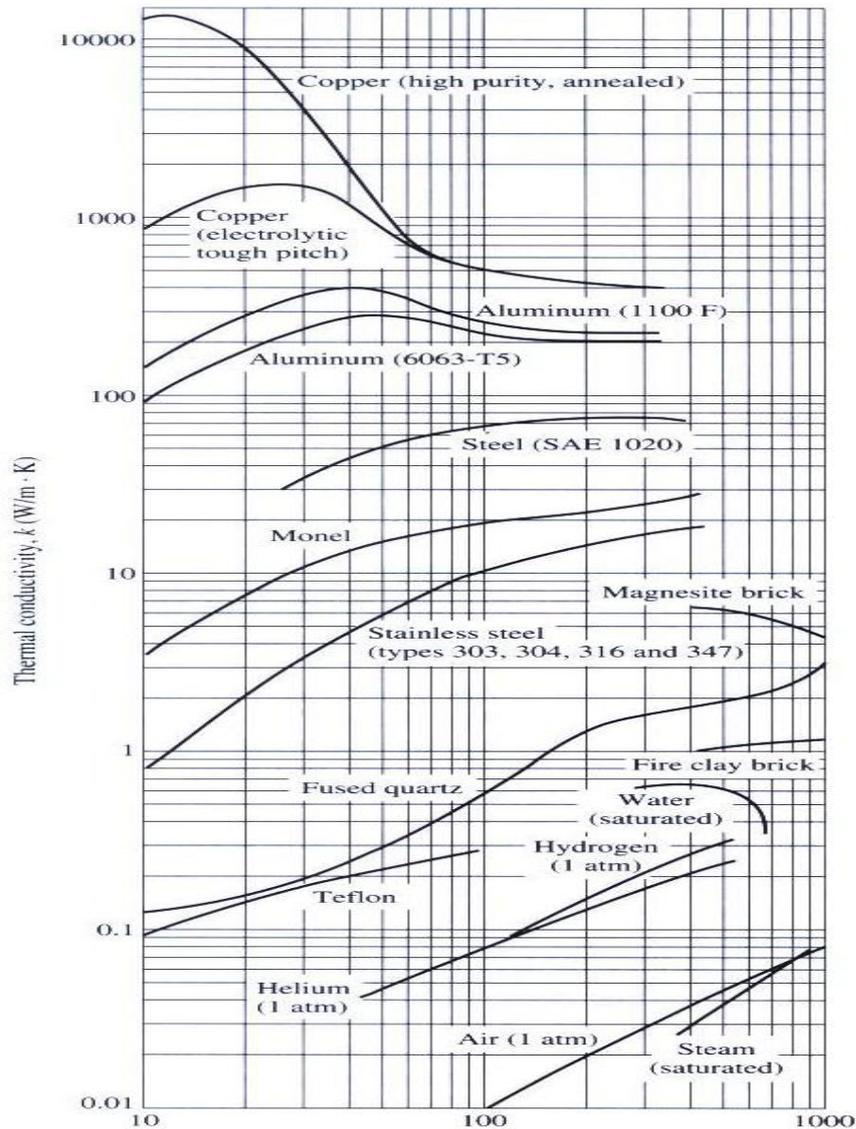

**Fig. 6:** Thermal conductivity (W m$^{-1}$ K$^{-1}$) of selected materials as a function of temperature (K). (Ref. [12].)

which states that the ratio between these two properties is, in good approximation, a constant for all metals at the same temperature. The proportionality constant $L$, known as the Lorenz number, is equal to:

$$L = 2.44 \times 10^{-8} \ (\text{W} \, \Omega \, \text{K}^{-2}) \,. \tag{4}$$

Electrical resistivity measurements, which are easier to make, therefore offer a more practical but indirect measure of thermal conductivity.

Let us now reconsider Fourier's law, but in one dimension, the case of a beam of cross-section $A$, and length $l$. Equation (1) in its integral form becomes (dropping the minus sign and keeping in mind that heat flows from high to low temperature):

$$\dot{Q} = \frac{A}{l} \cdot \int_{T_1}^{T_2} k(T) \, \mathrm{d}T \tag{5}$$

which expresses the conduction heat flow when the two tips of the beam are kept at temperatures $T_1$ and $T_2$. We can therefore note that the integral part is a property of the material in use and of the temperature range. For engineering applications, the use of tables of conductivity integrals is often practical. Table 3 gives an example for some of the most commonly used materials in cryostats:

**Table 3:** Thermal conductivity integrals (W m$^{-1}$) for selected technical materials (between indicated temperatures and 4.2 K).

|                                         | 20 K   | 80 K   | 290 K   |
|-----------------------------------------|--------|--------|---------|
| OFHC copper                             | 11,000 | 60,600 | 152,000 |
| DHP copper                              | 395    | 5,890  | 46,100  |
| Aluminium 1100                          | 2,740  | 23,300 | 72,100  |
| Aluminium 2024                          | 160    | 2,420  | 22,900  |
| Stainless steel AISI 304                | 16.3   | 349    | 3,060   |
| Typical glass-fibre/epoxy composite G-10 | 2      | 18     | 153     |

One can appreciate immediately that when designing a support for a superconducting device operating at 4.2 K, choosing a G-10 type glass-fiber epoxy composite can reduce conduction heat loads by a factor of about 20 as compared to steel (this is of course only an indicator in the choice of the material because other performance metrics like strength or stiffness have to be included).

Let us now consider once again the case of a beam of area *A* and length *L*, as illustrated in Fig. 8, where heat is transferred along *x*, and where an internal energy is deposited inside its volume.

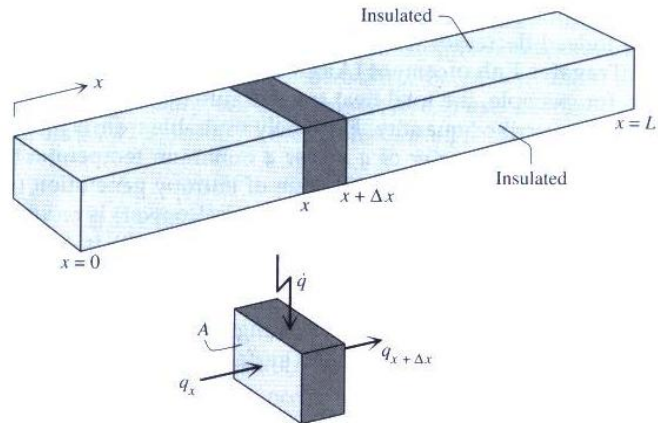

**Fig. 7:** One-dimension solid beam, with heat deposition *q* and conduction heat flux along *x*

We can write the energy balance equation for an element $x + \Delta x$, introduce Fourier's law describing longitudinal heat transfer, and introduce the change of internal energy related to the heat capacity of the element as:

$$\frac{\partial}{\partial x}\left(k\cdot\frac{\partial T}{\partial x}\right)+\dot{q}=\rho c\cdot\frac{\partial T}{\partial t} \tag{6}$$

where the first term represents longitudinal conduction, $\dot{q}$ is the internal heat deposited, and the term on the right-hand side represents the thermal inertia of the element related to its heat capacity ($\rho$ is the density and *c* is the specific heat capacity).

If *k* is considered constant (valid for small changes of *T*), we can rewrite the equation as:

$$\frac{\partial^2 T}{\partial x^2}+\frac{\dot{q}}{k}=\frac{1}{\alpha}\cdot\frac{\partial T}{\partial t} \text{ with } \alpha=\frac{k}{\rho c}. \tag{7}$$

Thermal diffusivity $\alpha$ indicates how fast a thermal perturbation develops along the beam. A few cases of special interest are detailed below.

### 3.1.1 Steady-state beam with no heating

Of particular interest is the steady-state case $(\partial T / \partial t = o)$ in which there is no internal heat deposition $(\dot{q} = 0)$. This is, for example, the case for any solid element having its extremities fixed in temperature ($T_0$ and $T_L$). Equation (6) is then simply re-written:

$$\frac{\partial}{\partial x}\left( k(T) \cdot \frac{\partial T}{\partial x} \right) = 0 \, . \tag{8}$$

We can distinguish two cases of interest:

(i) $k$ is constant. By integrating and imposing the boundary conditions at the extremities, we find a linear temperature profile along the beam, given by:

$$T = T_0 + \frac{(T_L - T_0)}{L} \cdot x \tag{9}$$

and a constant heat flux along the beam, given by:

$$\dot{q} = k \cdot \frac{A}{L}(T_L - T_0) \, ; \tag{10}$$

(ii) $k$ is linear with $T$ (in the example of impure metals, like steels and Al alloys). Let us assume $k = T/a$. By integrating and imposing the boundary conditions at the tips, we find a quadratic temperature profile along the beam, given by:

$$T = \sqrt{T_0^2 + \frac{T_L^2 - T_0^2}{L} \cdot x} \tag{11}$$

and a constant heat flux along the beam, given by:

$$\dot{q} = \frac{1}{2} A / L \cdot \frac{(T_L^2 - T_0^2)}{a} \, . \tag{12}$$

These two solutions are graphically shown in Fig. 8.

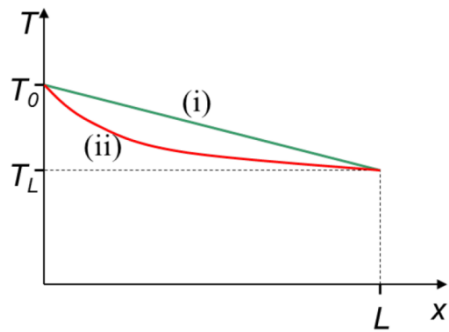

**Fig. 8:** Temperature profiles for solutions (i) and (ii)

### 3.1.2 Steady-state beam with uniform heat deposition

Another relevant case is that of a structure that is thermally connected to a heat sink at a given temperature, and which is subject to uniformly distributed heating (Fig. 9). This is the case, for example, of a thermal shield that is actively cooled by a cryogenic line at a fixed temperature at one point, and on which radiation heat is uniformly distributed and transferred through solid conduction from the hottest point to the cooling point.

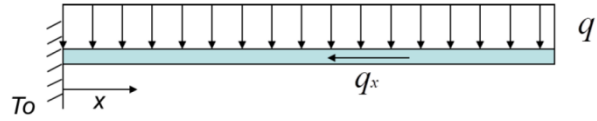

**Fig. 9:** Uniformly heated beam, cooled at one tip

Let $L$ be the beam length, $t$ its thickness, and $w$ its width (perpendicular to the page). The beam is cooled at $T_0$ in $x = 0$. The uniform heat deposition along the beam is $q$ (W m$^{-2}$).

Equation (6) can be re-written as:

$$\frac{d}{dx}\left(k\left(T\right)\cdot\frac{dT}{dx}\right)+\frac{q}{t}=0 \quad . \tag{13}$$

Considering $k$ to be constant (which is usually the case for thermal shields, as they should be designed to be quasi-isothermal), the equation can be integrated to yield the following temperature profile and heat flux along $x$:

$$T\left(x\right)=T_0-\frac{q}{2\cdot k\cdot T}\cdot x^2+\frac{q\cdot L}{k\cdot T}\cdot x \quad , \tag{14}$$

$$\dot{q}(x)=q\cdot w\left(x-L\right) \quad . \tag{15}$$

From Eq. (14), we can deduce the maximum $\Delta T$ along the beam, which may be the design objective achieved by providing a minimum thickness $t$. As previously mentioned, thermal shields are normally designed to be quasi-isothermal (typically within 5–10 K). So, by resolving $t$ as a function of $\Delta T_{\max}$ we obtain the practical design formula to choose the minimum thickness:

$$t\geq\frac{q\cdot L^2}{2k\cdot\Delta\cdot T_{\max}} \quad . \tag{16}$$

### 3.2 Residual gas conduction

Heat transfer by convection between surfaces at different temperature inside cryostats is negligible at low pressure, but residual gas conduction remains an important contribution that depends on the level of vacuum, the gas species, and the geometry and temperatures involved.

During normal operation, the large cryogenic cryo-pumping capacity of the cold surfaces of the internal components (essentially the surfaces of cryogen reservoirs and thermal shields) keeps pressure low enough (below $10^{-3}$ Pa) to provide an adequate insulation vacuum; vacuum pumping groups can also be stopped without observing any increase in pressure. Residual gases inside the cryostat or air leaks are condensed on the cold surfaces. Cryo-condensation depends on the gas species and on the temperature of the cold surfaces. More precisely, the condensation process happens when the gas pressure is higher than its vapour pressure at the temperature of the condensation surface. Figure 11 shows the vapour pressure curves for gases detected inside the LHC cryostats, which are representative of an industrial construction level of surface cleanliness (not Ultra High Vacuum, cleaned) and making use of MLI (mostly contaminated with water but also other types of hydrocarbon species). One can observe that at 4.2 K (liquid helium temperature), most of the gas species (with the exception of $H_2$ and He) can be cryo-condensed on the cold surfaces and the residual pressure always remains below $10^{-3}$ Pa. In the presence of He, having a vapour pressure well above the $10^{-3}$ Pa threshold even at 4.2 K or lower temperature, other pumping means must be employed to keep a good insulation vacuum (turbomolecular pumps, cryo-adsorbing materials, cryo-pumping in a molecular regime, etc.).

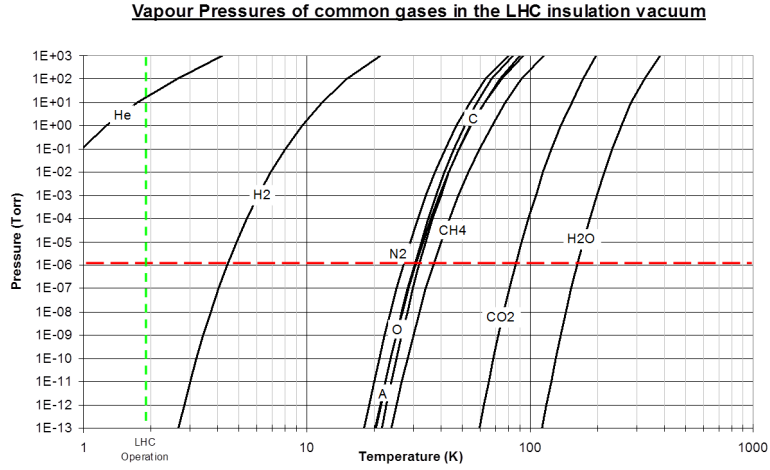

**Fig. 10:** Vapour pressures of common gases in the LHC insulation vacuum

Let us now consider heat transfer from residual gas conduction in the case of a degraded vacuum in cryostats. We are especially interested in the case of helium because practically all superconducting accelerator devices (magnets and accelerating cavities) are cooled in cryogenic helium, hence with the risk of helium leaks inside the cryostats.

Heat transmission between two surfaces separated by a gas follows two different regimes depending on the ratio between the mean free path of gas molecules $\lambda$ and the distance $L$ between the two surfaces. When $\lambda \ll L$, the *viscous regime* applies, and heat transmission is described in terms of thermal conduction $k$, and does not depend on pressure. In this case heat flux is inversely proportional to $L$ as in the case of solid conduction. Decreasing the residual gas pressure, the *molecular regime* is reached as $\lambda \gg L$. The molecules travel from the warm to the cold surface and heat transfer becomes proportional to residual gas pressure and is independent of wall distance $L$.

From gas kinetic theory, $\lambda$ is given by:

$$\lambda = 115 \frac{\eta}{p} \cdot \sqrt{\frac{T}{M}} \tag{17}$$

with viscosity $\eta$ in Pa.s, pressure $p$ in Pa, temperature $T$ in K, and molecular weight $M$ in g mol$^{-1}$.

In the molecular regime, residual gas conduction, from a warm surface $A_2$ at temperature $T_2$, to a cold surface $A_1$ at temperature $T_1$ obeys Kennard's law:

$$\dot{Q} = A1 \cdot \alpha(T) \cdot \Omega \cdot p \cdot (T_2 - T_1) \tag{18}$$

with pressure $p$ in Pa, temperatures in K, $\Omega$ is a coefficient depending on the gas species (2.13 W m$^{-2}$.Pa.K for helium), and $\alpha(T)$ is an *accommodation coefficient* that depends on the gas species, temperatures, and geometry of the surfaces. For simple geometries, as in the case of parallel or concentric walls, it is given by the expression:

$$\alpha = \frac{\alpha_1 \cdot \alpha_2}{\alpha_2 + \alpha_1 \left(1 - \alpha_2\right) \cdot \dfrac{A_1}{A_2}} \tag{19}$$

where $\alpha_1$ and $\alpha_2$, for helium and air at various temperatures, is given in Table 4:

**Table 4:** Accommodation coefficients for helium and air

| $T$ (K) | Helium | Air |
|---------|--------|-----|
| 300 | 0.3 | 0.8 |
| 80 | 0.4 | 1 |
| 20 | 0.6 | - |
| 4 | 1 | - |

We can note that residual gas conduction is proportional to pressure and to the temperature difference between walls.

### 3.3 Radiation

Thermal radiation is by far the most important contribution to the thermal budgets for cryostats, and deserves a special attention.

The surface of a body at a given temperature emits and absorbs electromagnetic radiation. When receiving electromagnetic radiation, a body absorbs a fraction $\alpha$ (absorptivity), reflects a fraction $\beta$ (reflectivity), and transmits a fraction $\tau$ (transmissivity), with energy conservation imposing the condition $\alpha + \beta + \tau = 1$. When $\tau = 0$, the body is said to be opaque, meaning that all energy is either absorbed or reflected. Also, when $\beta = 0$, and from energy conservation $\alpha = 1$, then the body is said to be *black*. In this case, the body absorbs all energy.

In general, the *opaque body* is a good approximation in most of our applications in cryostats. Conversely, the *black body* approximation almost never applies but is often used as a convenient simplification for calculating the maximum emitting radiation power of a body, and for first-order heat load calculations. But there is one case in cryostats where a surface behaves like a black body: an orifice in a thermal shield, through which radiation shines, and through multiple reflections is gradually absorbed by the internal surfaces (Fig. 11). The cryostat engineer has to be very careful, when designing thermal shields, in avoiding gaps and slots, which may be detrimental to the thermal performance. When gaps are unavoidable, typically when thermal contraction compensation gaps must be introduced, practical solutions can be adopted to avoid light shining through: for example the use of MLI pads covering slots; or if MLI cannot be used (in UHV applications) gaps should be surrounded by traps of high-absorptivity materials (special coatings, for example) to absorb light on the thermal shield wall, and reduce multi-path reflection on the inside.

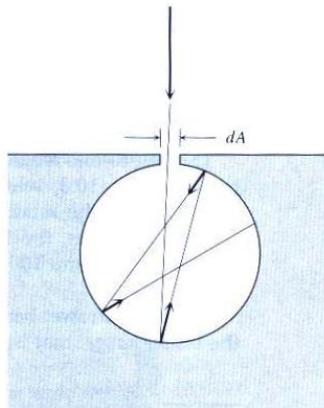

**Fig. 11:** Black body approximation of an aperture in a cavity

A black body emits a radiation flux with an emissive power, which is a function of the wavelength and depends on the temperature of the body, as illustrated in Fig. 12. It is interesting to note that the

maximum emissive power occurs at a wavelength inversely proportional to temperature, according to Wien's law:

$$\lambda_{max} = \frac{2898}{T} \quad (\mu \text{ m K}^{-1}) \tag{20}$$

and that at temperatures of practical interest in cryostats (from 300 K to vanishingly low temperatures), the maximum occurs at wavelengths in the far-infrared region.

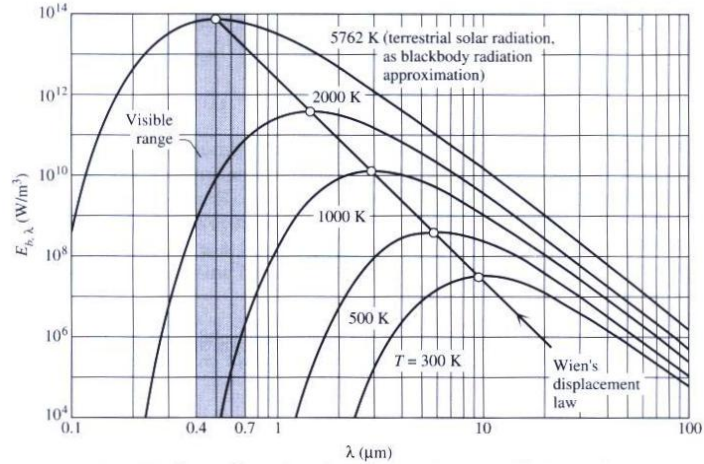

**Fig. 12:** Emissive power of a black body

By integrating over the whole wavelength spectrum, we can calculate the total emissive power:

$$E_b(T) = \int\limits_0^\infty E_{b,\lambda}\, d\lambda = \sigma \cdot T^4 \quad (\text{W m}^{-2}) \tag{21}$$

with $\sigma$ being the Stefan-Boltzmann's constant:

$$\sigma = 5.67 \times 10^{-8} \quad (\text{W m}^2 \text{ K}^4) \tag{22}$$

providing the important result that the total emissive power is therefore proportional to the fourth power of temperature. From simple calculation, we can conclude that a black body at room temperature (293 K) emits about 420 W m$^{-2}$. Considering that the vacuum vessel of a cryostat is at room temperature, and assuming that a thermal shield has a 1-cm$^2$ gap, results in a black body heat load through the gap of ~10 mW. The value may appear small, but when compared to the extremely high electrical power cost of heat extraction at cryogenic temperatures (about 1 kW W$^{-1}$ from 1.8–300 K), 10 mW deposited at 1.9 K requires about 10 W of cooling power, and is therefore three orders of magnitude larger.

Let's now consider heat exchange between two black bodies at different temperature and facing each other with a given orientation, as illustrated in Fig. 13.

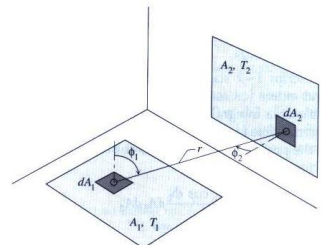

**Fig. 13:** Scheme of two surfaces exchanging radiation heat

The heat exchange balance between the two can be computed as the difference between radiation emitted from the first body and absorbed by the second, and the radiation emitted by the second body and absorbed by the first. This is given by:

$$q_{1-2} = \sigma(T_1^4 - T_2^4)A_1F_{12} \tag{23}$$

where the *geometrical view factor* $F_{12}$ is the fraction of the total radiation leaving the first body, which is intercepted and absorbed by the second and which depends upon the relative orientation of the two surfaces. It can be demonstrated that reciprocity applies, and that $A_1F_{12} = A_2F_{21}$.

View factors are quite tedious to calculate, but values can be found in the literature. Figure 14 gives view factors for selected geometries.

| Configuration | Geometric View Factor |
|---|---|
| 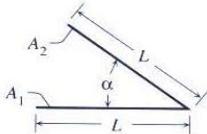 | Two infinitely long plates of width $L$, joined along one of the long edges:<br><br>$$F_{12} = F_{21} = 1 - \sin\frac{\alpha}{2}$$ |
| 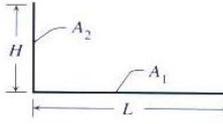 | Two infinitely long plates of different widths $(H, L)$, joined along one of the long edges and with a $90°$ angle between them:<br><br>$$F_{12} = \tfrac{1}{2}[1 + x - (1 + x^2)^{1/2}]$$<br>where $x = H/L$ |
| 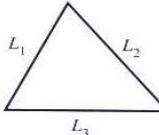 | Triangular cross section enclosure formed by three infinitely long plates of different widths $(L_1, L_2, L_3)$:<br><br>$$F_{12} = \frac{L_1 + L_2 - L_3}{2L_1}$$ |
| 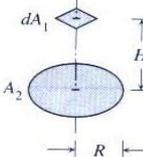 | Disc and parallel infinitesimal area positioned on the disc centerline:<br><br>$$F_{12} = \frac{R^2}{H^2 + R^2}$$ |
| 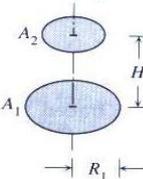 | Parallel discs positioned on the same centerline:<br><br>$$F_{12} = \frac{1}{2}\left\{X - \left[X^2 - 4\left(\frac{x_2}{x_1}\right)^2\right]^{1/2}\right\}$$<br>where $x_1 = \frac{R_1}{H}$, $x_2 = \frac{R_2}{H}$, and $X = 1 + \frac{1+x_2^2}{x_1^2}$ |
| 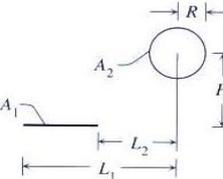 | Infinite cylinder parallel to an infinite plate of finite width $(L_1 - L_2)$:<br><br>$$F_{12} = \frac{R}{L_1 - L_2}\left(\tan^{-1}\frac{L_1}{H} - \tan^{-1}\frac{L_2}{H}\right)$$ |
| 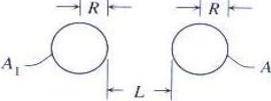 | Two parallel and infinite cylinders:<br><br>$$F_{12} = F_{21} = \frac{1}{\pi}\left[\left(X^2 - 1\right)^{1/2} + \sin^{-1}\left(\frac{1}{X}\right) - X\right]$$<br>where $X = 1 + \frac{L}{2R}$ |

**Fig. 14:** View factors for given geometries

Real surfaces do not behave as black bodies but rather as *grey-diffuse bodies*. Only a fraction of the black body radiation is emitted (i.e. *grey-body*), and the intensity of emission is uniform in all directions, (i.e. *diffuse-emitting body*).

By introducing the total hemispheric *emissivity,* which defines this fraction of the emitted radiation with respect to that of a black body:

$$\varepsilon(T) = \frac{E(T)}{E_b(T)} \le 1. \tag{24}$$

Equations (21) and (23) become, respectively:

$$E(T) = \varepsilon \cdot \sigma \cdot T^4 \quad, \tag{25}$$

$$q_{1-2} = \varepsilon \cdot \sigma (T_1^4 - T_2^4) A_1 F_{12}. \tag{26}$$

For real materials, the grey-diffuse model is a satisfactory approximation. Emissivity depends on the material but also on the surface finish and cleanliness. Clean and well-polished metallic surfaces have small emissivity, whereas non-metallic surfaces have higher emissivity. For this reason thermal shields in cryostats are normally made of metallic surfaces (copper or aluminium), and with a clean and reflective surface finish. Table 5 gives hemispherical emissivity values for a selection of materials.

**Table 5:** Hemispheric emissivity for a selection of materials

| | Temperature (K) | | | |
|---|---|---|---|---|
| | 4 | 20 | 80 | 300 |
| Copper mechanically polished | 0.02 | | 0.06 | 0.1 |
| Copper black oxidised | | | | 0.8 |
| Gold | | | 0.01 | 0.02 |
| Silver | 0.005 | | 0.01 | 0.02 |
| Aluminium electropolished | 0.04 | | 0.08 | 0.15 |
| Aluminium mechanically polished | 0.06 | | 0.1 | 0.2 |
| Aluminium with 7 μm oxide | | | | 0.75 |
| Magnesium | | | | 0.07 |
| Chromium | | | 0.08 | 0.08 |
| Nickel | | | 0.022 | 0.04 |
| Rhodium | | | 0.08 | |
| Lead | 0.012 | | 0.036 | 0.05 |
| Tin | 0.012 | | 0.013 | 0.05 |
| Zinc | | | 0.026 | 0.05 |
| Brass, polished | 0.018 | | 0.029 | 0.035 |
| Stainless steel, 18-8 | 0.2 | | 0.12 | 0.2 |
| Glass | | | | 0.94 |
| Ice | | | | 0.96 |
| Oil paints, any colour | | | | 0.92–0.96 |
| Silver plate on copper | | 0.013 | 0.017 | |
| Aluminium film 400A on Mylar | | | 0.009 | 0.025 |
| Aluminium 200A on Mylar | | | 0.015 | 0.035 |
| Nickel coating on copper | | 0.027 | 0.033 | |

It should be noted that our appreciation of the visual appearance of the surface finish is limited to the visible range, and this may be misleading. For instance, a reflective metallic surface, which has been coated by a transparent resin, still appears reflective. But since the resin behaves like a black body at room temperatures or lower, it will absorb radiation rather than reflect it. This is the case of some aluminized tapes available on the market and commercialized for their low-emissivity properties of practical use in cryostats (for shielding parts, or joining MLI blankets), and in which the aluminium reflective layer is deposited onto a transparent polyester tape, therefore behaving like a black body despite the shiny appearance.

Emissivity varies with temperature. For metals at cryogenic temperatures emissivity reduces almost proportionally with temperature, which enhances the low-emissivity properties of cryogenic cooled thermal shields in cryostats.

Of particular interest to cryostat design are the geometries with two enclosed envelopes, in particular with cylinders, and parallel flat plates.

For enclosed cylinders, with $A_1$ and $A_2$ being the inner and outer surfaces respectively, and kept at temperatures $T_1 < T_2$, from energy balance and making use of Eq. (26), the power exchanged is given by:

$$q_{1-2} = \frac{\sigma A_1 (T_2{}^4 - T_1{}^4)}{\dfrac{1}{\varepsilon_1} + \dfrac{A_1}{A_2}\left(\dfrac{1}{\varepsilon_2} - 1\right)} F_{12} \; . \tag{27}$$

This equation provides interesting design hints. Assuming $A_2$ is the vacuum vessel of a cryostat and $A_1$ is a given internal cryogenic component (or a thermal shield), reducing radiation heat loads to $A_1$ can be obtained by reducing the size of the vessel $A_2$ (the radiation source), as well as reducing both $\varepsilon_1$ and $\varepsilon_2$ (in particular $\varepsilon_1$, which has the largest effect). For this reason thermal shields should be made with low-emissivity surfaces (aluminium, copper or Ni plating), whereas the vacuum vessel, dominated by other constraints (construction materials and fabrication technology) is normally not prepared for low emissivity.

For parallel flat plates of area $A$, which is also a convenient approximation for surfaces of large radius placed together at a close distance, the power exchange, when $T_2 < T_1$, is given by:

$$q_{1-2} = \frac{\sigma A (T_1{}^4 - T_2{}^4)}{\dfrac{1}{\varepsilon_1} + \dfrac{1}{\varepsilon_2} - 1} \; . \tag{28}$$

If we now insert a third intermediate floating screen between the two flat surfaces, and we assume that $\varepsilon_1 = \varepsilon_2 = \varepsilon$, a good approximation when materials are similar and their temperatures are not too different, the heat exchange becomes:

$$q_{1-2} = \frac{\sigma (T_1{}^4 - T_2{}^4)}{2\left(\dfrac{2}{\varepsilon} - 1\right)} \; , \tag{29}$$

which is therefore reduced to half of the thermal radiation between the two parallel surfaces.

This is a useful hint on how radiation heat loads in a cryostat can be reduced by inserting one (or more) floating reflective screens between the vacuum vessel and the thermal shield. This introduces the topics of thermal shielding and MLI, presented in the following sections.

### 3.4 Thermal shielding

Introducing one intermediate floating shield in a cryostat reduces by half the heat load to the cold surface. If a dedicated cryogenic circuit actively cools the shield, its effectiveness can be improved by two mechanisms. First, the emissivity of the shield decreases at lower temperatures, therefore the heat exchanged with the vacuum vessel is reduced as well as the heat radiation between the thermal shield and the cold surface. Secondly, heat extraction by active cooling of the thermal shield at a higher temperature than that of the cold surface results in a lower cryogenic cooling cost (as explained in Section 4).

Thermal shielding cooled at a temperature in the range of 50–80 K is a universal practice. In the LHC cryostats (Fig. 5), the thermal shield is made of aluminium, and is supported on the magnet support posts, for which it also provides heat interception. The shield is composed of an aluminium bottom tray extrusion, integrating a cooling line at 50–80 K, which provides mechanical stiffness to 15-m long assemblies, and is closed on its top part by a rolled and welded 2.5-mm thin aluminium sheet. Given the very large quantity of material required, aluminium was preferred to copper for its lower cost while still offering a reasonably good thermal conductivity.

Thermal radiation protection of the thermal shields is enhanced by the use of MLI, discussed in the next section.

### 3.5    Multilayer insulation protection

MLI is based on the principle of multiple radiation reflection obtained by inserting reflective layers between the warmer radiating surface (typically the vacuum vessel) and the colder surfaces (thermal shield or cold surface of the internal device).

The result of Eq. (29) can be generalized to $N$ reflective layers, obtaining a reduction of radiation by a factor $(N+1)$. In practice, reflective layers are packed in blankets, and thermal contact between adjacent layers would be inevitable; for this reason, the reflective layers are interleaved with insulating spacers, as schematically shown in Fig. 15, limiting thermal conduction.

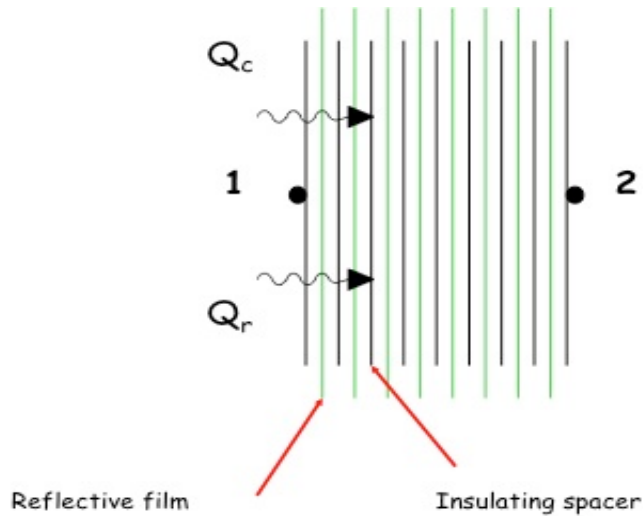

**Fig. 15:** MLI structure, showing reflective layers interleaved with insulating spacers

Modelling the thermal heat exchange through MLI can be quite complex, material (reflective and insulating layers) dependent, as well as strongly influenced by the application of the blanket in the cryostats. A simplified engineering model considers two main contributions, one accounting for radiation, proportional to the difference of temperatures at the fourth power, and a second accounting for the residual solid conduction across the blanket thickness proportional to the temperature difference between the warm and the cold surface and to the average temperature:

$$q_{\mathrm{MLI}} = \left[ \frac{\beta}{N+1} \cdot (T_1^4 - T_2^4) \right] + \frac{\alpha}{N+1} \cdot \frac{T_1 + T_2}{2} \cdot (T_1 - T_2) \qquad (30)$$

where $\beta$ and $\alpha$ are corrective factors that should be obtained experimentally. Values obtained for the MLI system of the thermal shield of the LHC cryostats are $\beta = 3.741 \; 10^{-9}$ and $\alpha = 1.401 \; 10^{-4}$.

MLI in accelerator cryostats generally consists of aluminised thin polyethylene films as reflective surfaces, and paper, glass-fibre, or polyester insulating nets as insulating spacers.

The efficiency of MLI also depends on the practical aspects of how it is mounted. The packing density of the layers plays an important effect. Figure 16 shows schematically that with increasing packing density, the solid conduction contribution increases. There is a somewhat optimal packing density, which can be found between 15 and 25 layers per centimetre.

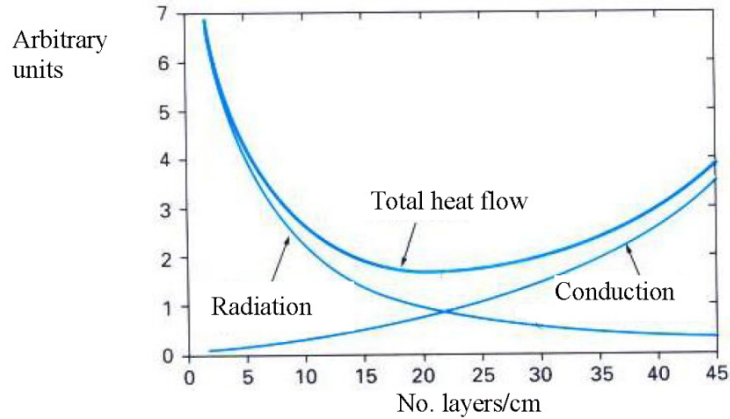

**Fig. 16:** Total heat flow as a function of MLI packing density

MLI efficiency also depends on practical implementation rules. The ideal packing density should be preserved after installation of the blankets. When mounted in horizontal accelerator cryostats, MLIs tend to be compressed by their weight on the top part of the thermal shields, whereas it tends to be looser at the bottom. A good rule is to check that an acceptably correct packing remains after installation. Also, when preassembled MLI blankets are to be mounted around circular geometries, one should include in the design the larger circumferential dimension of the outermost layers with respect to innermost ones. Differential thermal contractions between the MLI and its support should be also considered to preserve the correct packing when cold. Finally, a typical mistake is to stack too many layers in a reduced space between warm and cold surfaces, ending up with unwanted thermal conduction shorts. A good rule is to leave a minimum gap of at least a few centimetres between the warm surface and the outermost MLI layer.

The choice of the *ideal* number of MLI layers is not a straightforward one, and is very much dependent on the materials employed (film material and thickness, single or double-aluminisation, aluminisation thickness, type of spacers, etc.); a general tendency is to introduce as many layers as can possibly fit in the available space, sometimes resulting in thermal contact shorts, but also in an increase of material and assembly costs.

Extensive testing was carried out at CERN on MLI types and blanket assemblies for the LHC magnet cryostats. An optimal and cost-effective solution was chosen for the MLI blankets of the thermal shield (Fig. 17): 30 reflective layers (double-aluminised polyethylene terephthalate film, coated with a minimum of 400 Å of aluminium on each side) interleaved with polyester net. An additional ten-layer MLI blanket is mounted on the cold mass, essentially to reduce heat loads from residual helium gas conduction in case of degraded vacuum from helium leaks, as well as thermally insulating the cold surfaces in case of an accidental air venting of the cryostat.

MLI installation precautions were carefully studied for the LHC cryostats, also including industrial production and assembly considerations to be compatible with the large series (about 2000 dipole and quadrupole cryostats).

MLI blankets were preassembled in industry, and assembled in the magnet cryostats at CERN (Fig. 17). An important investment in tooling for series production of blankets allowed the manufacturer to reduce manpower costs.

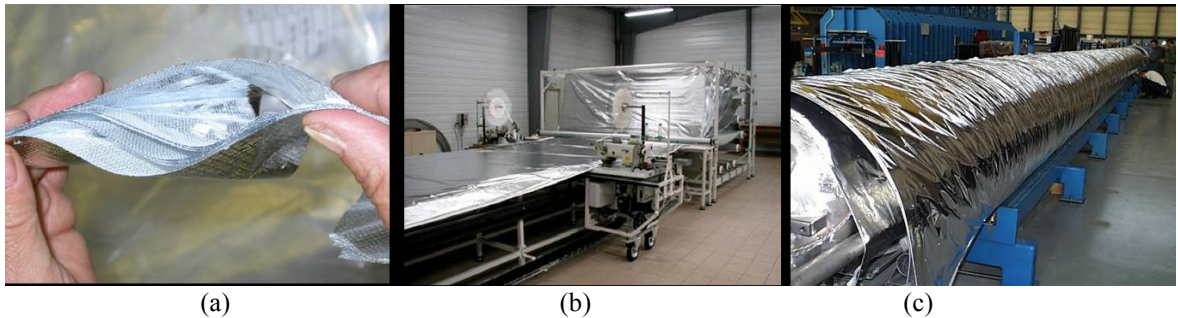

<div align="center">(a)        (b)        (c)</div>

**Fig. 17:** (a) MLI stack in a blanket; (b) stitching table for industrial production; (c) MLI blanket on a dipole cold mass and thermal shield.

The thermal performance of the LHC cryostat system was experimentally measured at various stages of the design and construction of the machine, but final thermal performance measurements were made during commissioning of the machine.

From these results the performance of the MLI system could be deduced, yielding the following figures:

- Thermal efficiency of 30 layers of LHC MLI between 300 K and 50 K: ~1 W m$^{-2}$;

- Thermal efficiency of 10 layers of LHC MLI between 50 K and 1.9 K: ~50 mW m$^{-2}$.

The literature reports laboratory tests providing lower values than these, but the interest of the above figures is that they include all the inefficiencies resulting from the real implementation of MLI protection on the large-scale accelerator.

To conclude this first part dedicated to thermal studies of the cryostats, the reader can find in Appendix A a case study based on the LHC cryostat with a numerical application of the formulae presented on radiation protection, thermal shielding, MLI, and residual gas conduction. Appendix B presents the experimental validation of the thermal performance of the LHC cryostats during the commissioning of the machine.

## 4    Cryogenics considerations

Helium is the cryogenic fluid of choice in accelerator technology because its thermo-physical properties make it the only one suitable to operate superconducting devices below 10 K. When operating below the *lambda point*, at temperatures below 2.17 K, helium offers the possibility of enhancing the field performance of magnets made in Nb-Ti, while superconducting RF acceleration cavities can profit from a reduction of the Bardeen, Cooper and Schrieffer (BCS) contribution to the surface resistance with a reduction in dynamic losses, especially at high frequency.

Accelerator magnets are usually operated with sub-cooled liquid (i.e. pressurized above its saturation point) ensuring a good liquid coverage and penetration around the superconducting coils. Operating in pressurised liquid is also beneficial in reducing the risk of electrical insulation breakdown in gaseous helium and avoids contamination of helium in the case of leaks. Superconducting Radio Frequency (SRF) cavities are normally operated in saturated helium, where pool-boiling cooling is preferred for technical simplicity of the cooling schemes.

Figure 18 illustrates on a phase diagram the helium operating point of a number of state-of-the-art linear and circular superconducting accelerators.

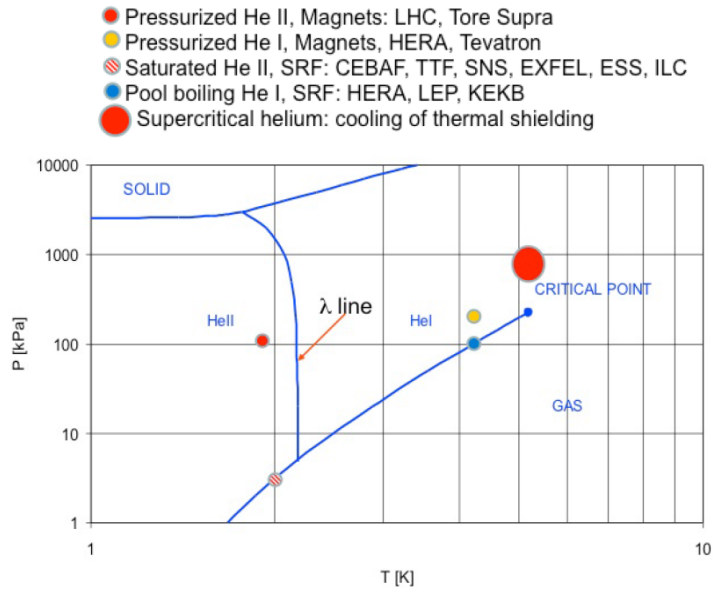

**Fig. 18:** He phase diagram and operation point of several superconducting machines

Cooling at higher temperatures than the operating point of the superconducting device is needed in cryostats for thermal shields and intermediate temperature heat interception in cryostat components. A temperature of about 80 K is often chosen, with the advantage that liquid nitrogen can be employed, a far more cost-effective cryogenic fluid.

From Table 6, comparing the main thermo-physical properties of helium and nitrogen, we note that the latent heat of evaporation of nitrogen is one order of magnitude larger than that of helium, meaning that for a given heat load the boil-off mass flow is ten times lower than for helium, making it a more effective coolant when large enthalpies have to be extracted, typically when cooling down massive equipment from ambient temperature down to 80 K. This is the case for the initial cool-down phase of the LHC magnets, where the helium circulated in the magnets is pre-cooled by liquid nitrogen. On the contrary, helium vapour has seven times higher enthalpy between its boiling temperature and 300 K (1550 kJ kg$^{-1}$), meaning that its vapours still have a large cooling capacity that can be effectively used for cooling components with a thermal conduction path from room temperature to 4.2 K, as explained in the following sections.

**Table 6:** Main thermo-physical properties of helium and nitrogen

| Property | Units | 4He | N$_2$ |
|---|---|---|---|
| Boiling $T$ (at 1 atm) | K | 4.2 | 77.3 |
| Critical temperature | K | 5.2 | 126.1 |
| Critical pressure | 105 Pa | 2.23 | 33.1 |
| Latent heat of evaporation (at 1 atm) | kJ kg$^{-1}$ | 21 | 199 |
| Enthalpy between $T$ boiling and 300 K | kJ kg$^{-1}$ | 1550 | 233 |
| Liquid density (boiling at 1 atm) | kg m$^{-3}$ | 125 | 810 |
| Saturated vapour density (at 1 atm) | kg m$^{-3}$ | 17 | 4.5 |
| Gas density (at 1 atm, 273.15 K) | kg m$^{-3}$ | 0.18 | 1.25 |
| Liquid viscosity (at boiling $T$) | μPa s | 20 | 17 |

### 4.1 Efficiencies of helium refrigeration and liquefaction

Heat loads are extracted from the cryostat either through boil-off of the liquid cryogens and pumping of vapours, or through forced circulation of cryogens (typically in their supercritical state) in cooling circuits with heat exchangers. In the first case, isothermal cooling can be maintained by ensuring constant pressure by vapour pumping over the saturated bath. This is normally adopted to keep the constant operating temperature of the superconducting device. The second case provides non-isothermal cooling and is generally the way that thermal shields and heat intercepts are cooled; a temperature increase along the cooling line is inevitable but the cooling circuit can be designed to keep it conveniently small.

In both cases the heat extracted from the cryostats is transported, through a mass flow of cryogens, to the refrigerator, where heat is eventually rejected to the ambient. The cryogen is then recompressed, cooled, and circulated back to the cryostat and its utilities.

In the ideal case, refrigeration requires a minimum of work, according to Carnot's principle, to extract a given heat from a temperature $T_c$ to a heat sink at $T_w$; in the $T$-$S$ diagram this corresponds to the area contained between two adiabatic and two isothermal transformations of a reversible cycle (Fig. 19.)

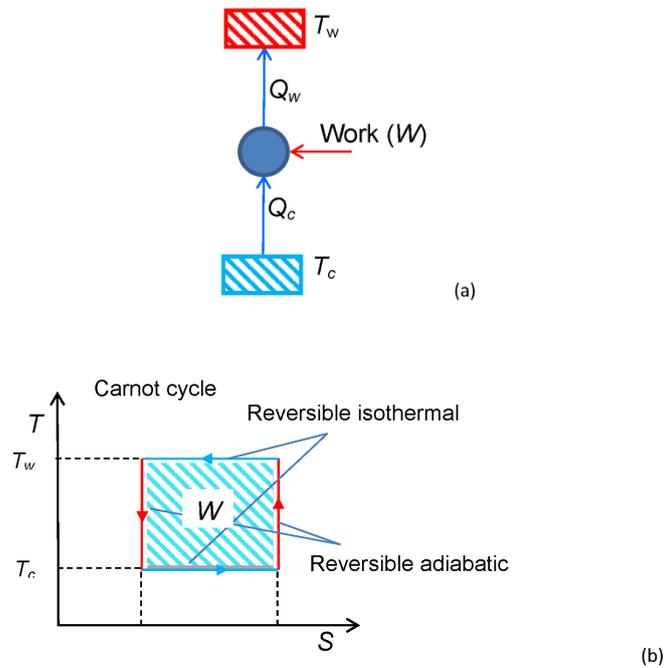

**Fig. 19**: (a) Refrigeration between a cold source Tc and a warm one Tw; (b) the ideal Carnot cycle

A corresponding coefficient of performance (*COP*) can be defined as the ratio between the extracted heat from the cold source and the associated extraction work:

$$COP = \frac{Qc}{W} \ .$$ 

(31)

In the ideal case of the Carnot cycle, making use of the first and second principles of thermodynamics, the COP attains its maximum value:

$$COP_{max} = \frac{T_c}{T_w - T_c}$$

(32)

simply depending on temperatures or, by combining Eqs. (31) and (32), calculate the minimum work needed to extract a given heat $Qc$:

$$W_{min} = Qc \cdot \frac{T_w - T_c}{T_c} \quad . \tag{33}$$

Refrigerators eventually dump heat to ambient temperature, so by taking $T_w = 293$ K, we can easily calculate the minimum specific heat extraction work, also equivalent to the specific heat load extraction power in W/W, for a refrigerator working between the liquid temperature of commonly used cryogens and room temperature (Table 7).

**Table 7:** Minimum refrigeration power at different liquid temperatures

| Fluid | Temperature (K) | $W_{min}\,Qc^{-1}$ (W/W) |
|---|---|---|
| Liquid N$_2$ | 77 | 2.8 |
| Liquid H$_2$ | 20.4 | 13.4 |
| Liquid He | 4.2 | 68.4 |
| Liquid He, (operating T of LHC magnets) | 1.8 | 161.8 |

Extracting heat at 2 K rather than at 80 K implies a cost increase by a factor of more than 50. This is the reason why heat loads to the lowest temperature in a cryostat (the operating temperature of the superconducting device) have to be minimized, and why heat interception is thermodynamically more convenient at higher temperatures, which explains why thermal shields are cooled at higher temperatures, in the 50–80 K range.

In the real case, refrigerators have an efficiency even lower than the Carnot efficiency due to irreversibilities in the machine. Technical efficiency of refrigerators is only a fraction of the Carnot factor, approaching 0.3 for modern large refrigeration systems at 4.2 K. Therefore the net refrigeration efficiency at 4.2 K is only 0.0028; in other words cooling 1 W at 4.2 K costs an extraction power of about 360 W. In the case of the LHC magnets, which are operated at 1.8 K, where the efficiency is about 0.15 of Carnot, 1 W of heat load costs about 1 kW of extraction power (Fig. 20).

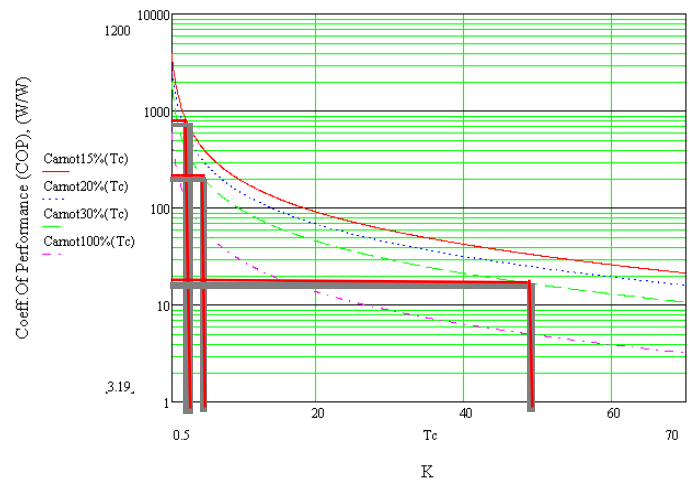

**Fig. 20:** *COP* as a function of cooling temperature (K) and percentage Carnot efficiencies

In cryogenic plants, a distinction can be made between operation in *pure refrigeration*, where isothermal cooling of the utilities load is provided, and *pure liquefaction*, in which liquid is produced and used for non-isothermal cooling of the utilities load, as illustrated in Fig. 21. Accelerator cryogenic plants generally provide a combination between these operating modes, depending on the specific needs, with some possibility of trading between the two.

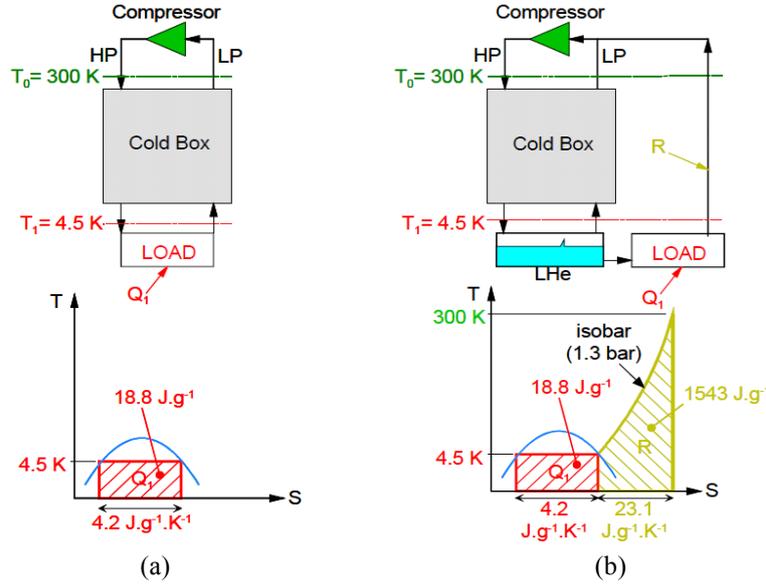

**Fig. 21:** (a) Simplified refrigeration versus (b) liquefaction cycles

When operating in liquefaction mode, the 4.5 K liquid produced can be used for isothermal boil-off cooling making use of its latent heat of evaporation, followed by non-isothermal cooling up to room temperature, making use of the large remaining cooling capacity of its vapours. Room temperature gaseous helium is then transferred back to the cryogenic plant for recompression and re-liquefaction. As schematically shown in the *T–S* plots in Fig. 21, the cooling capacity in the liquefaction mode is about 75 times higher than in refrigeration.

Let us now compare the thermodynamic efficiency of liquefaction and refrigeration at 4.2 K. By combining the first and second principles of thermodynamics, and introducing entropy $S$, we can express the work of refrigeration as:

$$W = T_w \cdot \Delta S_c - Q_c \ . \tag{33}$$

For liquefaction, we can split Eq. (33) into two contributions, the first one for pre-cooling helium from 293 K down to 4.5 K, and the second one for condensation; furthermore, considering that heat exchanged at constant pressure is equivalent to enthalpy $H$, we can write:

$$\begin{aligned} W_{liq} &= T_w \cdot \Delta S_{precool} - Q_{precool} + T_w \cdot \Delta S_{conden.} - Q_{conden.} \\ &= T_w \cdot \Delta S_{precool} - \Delta H_{precool} + T_w \cdot \Delta S_{conden.} - \Delta H_{conden.} \end{aligned} \tag{34}$$

From tables of thermodynamic properties of helium, we can extract the entropy and enthalpy figures and calculate the work of liquefaction, which turns out to be 6.6 kW *per* 1 g s$^{-1}$ of liquefied gas.

By taking the Carnot minimum specific heat refrigeration work of about 68 W/W, we can now conclude that 1 g s$^{-1}$ *of helium liquefaction is about equivalent to* 100 W *iso-thermal refrigeration at* 4.5 K. We now have the ingredients for comparing the thermal efficiency of vapour cooling and isothermal cooling of cryostat heat intercepts, presented in the following sections.

## 5    Heat intercepts

In the previous section we have seen that the cost of refrigeration increases dramatically when working at temperatures close to absolute zero. For this reason, the thermal design of cryostats, especially for systems operating with superfluid helium, should aim at containing heat loads reaching the lowest temperature level that may be achieved by thermal shielding and by heat interception. Thermal shielding was extensively covered in the previous sections, and the next sections are dedicated to heat interception.

### 5.1    Conduction heat intercepts

Solid conduction in cryostat components from room temperature to the lowest temperatures can be reduced by thermalizing at higher temperatures at intermediate and adequately chosen locations. The available intermediate temperature levels often depend on the architecture of the machine and its cryogenic system, but the use of a thermal shield cooled at a temperature in the range of 50–80 K is a universal practice, so this temperature level is also available for heat intercepts. In the LHC, a second cooling circuit at a temperature level of 5–20 K is also used to cool the beam screens inside the beam tubes; the same circuit is also used for a second level of heat interception on most components, in particular on the magnet support posts. See Fig. 22.

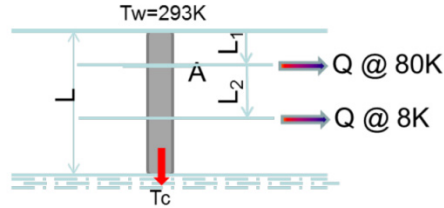

**Fig. 22:** Two heat intercepts at intermediate temperatures

By minimizing the following cooling cost function $f(L_1, L_2)$, sum of three conductive contributions weighted with three factors C1, C2, and C3, which depend on the cost of cryogenic cooling at the three temperature levels, we can find the optimal locations $L_1$, $L_2$ of the heat intercepts:

$$\min\left\{ f(L_1, L_2) = C1 \cdot \frac{A}{L_1} \int_{Tw}^{80K} k(T)\mathrm{d}T + C2 \cdot \frac{A}{L_2 - L_1} \int_{80K}^{8K} k(T)\mathrm{d}T + C3 \cdot \frac{A}{L - L_2} \int_{8K}^{T_c} k(T)\mathrm{d}T \right\} . \quad (35)$$

For the LHC support posts, with $T_c = 2$ K, the optimal positions of two heat intercepts, at 5 K and 75 K, were chosen to minimize the cost function with C1 = 16 W/W, C2 = 220 W/W, and C3 = 990 W/W as cost factors at 75 K, 5 K, and 1.8 K, respectively.

Table 8 compares the cryogenic cost for heat interception under various configurations. The two-heat intercepts solution chosen offers a cryogenic cooling cost saving of more than one order of magnitude as compared to that without heat intercepts.

**Table 8:** Heat loads with and without heat intercepts and total cost of heat extraction per LHC support post

| Number of heat intercepts (in optimal positions) | $Q_{1.8K}$ (W) | $Q_{5K}$ (W) | $Q_{75K}$ (W) | $Q_{elec}$ (W) |
|---|---|---|---|---|
| No intercept | 2.79 | - | - | 2790 |
| 1 (at 75 K) | 0.54 | - | 6.44 | 638 |
| 2 (at 75 K and at 5 K) | 0.047 | 0.42 | 7.1 | 252 |

Figure 23 illustrates an LHC support post in glass-fibre reinforced epoxy (GFRE) composite material with two aluminium heat intercepts positioned in optimal positions.

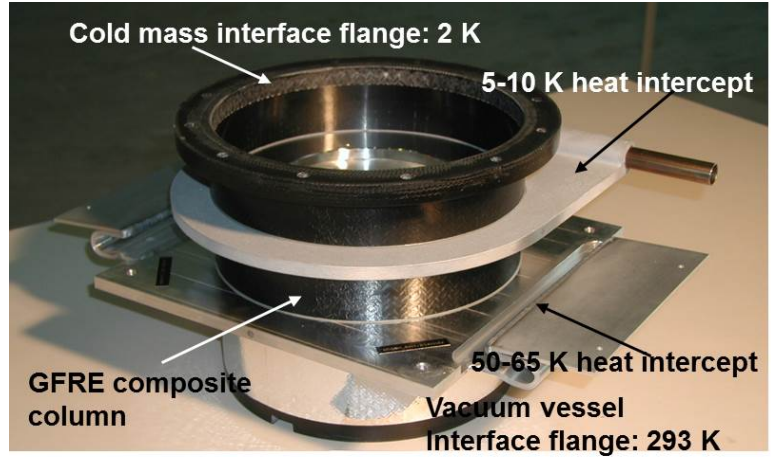

**Fig. 23:** LHC support post in GFRE, with two heat intercepts

The heat intercepts are 10-mm thick aluminium plates close-fitted and glued to the external wall of the composite column. The heat intercepts shrink-fit onto the composite column during cool-down ensuring an improved heat exchange due to contact pressure.

Active cooling of the bottom intercept at 50–65 K is ensured by an all-welded aluminium connection to the thermal shield, while the top intercept is a cast aluminium plate integrating a stainless steel cryogenic line at 5–10 K.

## 5.2    Vapour cooling

The large enthalpy stored in helium gases evaporating from a bath can be usefully employed for cooling supports, cryostat necks, RF couplers and current leads. Taking a support of length $L$ and cross-section $A$ (Fig. 24), considering perfect exchange with the escaping gas (i.e., assuming identical temperature of wall and gas), the heat balance equation gives:

$$k(T) \cdot A \cdot \frac{\mathrm{d}T}{\mathrm{d}x} = \overset{\bullet}{Q} + \overset{\bullet}{m} Cp \cdot \left( T - Tl \right) \tag{36}$$

where $\overset{\bullet}{Q}$ is the residual heat flow reaching the liquid bath at temperature $T_l$.

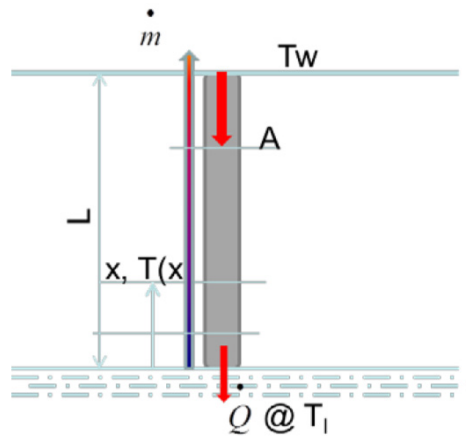

**Fig. 24:** Vapour cooled support with perfect cooling

Assuming that the vapour boil-off is entirely due to this heat load (so-called *self-sustained cooling*), and introducing the latent heat of evaporation $Lv$, we can state:

$$\overset{\bullet}{Q} = \overset{\bullet}{m} \cdot Lv \quad . \tag{37}$$

Equation (36) yields:

$$\overset{\bullet}{Q} = \frac{A}{L} \cdot \int_{T_l}^{T_w} \frac{k(T)}{1 + \dfrac{(T - T_l) \cdot Cp}{Lv}} \cdot \mathrm{d}T \quad . \tag{38}$$

In this form, the expression of the heat load is equivalent to that of solid conduction but for the denominator in the integral, which being larger than one results in an *attenuation factor* with respect to the conduction case.

Attenuation factors for liquid helium cooling can be in the range of 15 to 30 as reported in Table 9 for a selection of technical materials.

**Table 9:** Thermal conduction integrals and corresponding reduced thermal conductivity integral in self-sustained helium cooling from 4.2–293 K.

| Material | Thermal conductivity integral (W cm$^{-1}$) | Reduced thermal conductivity integral (W cm$^{-1}$) |
|---|---|---|
| OFHC copper | 1520 | 110 |
| Aluminium 1100 | 728 | 40 |
| AISI 300 stainless steel | 31 | 0.92 |

Vapour cooling is widely used for electrical conductors feeding current from room temperature to superconducting devices (Fig. 25). When designing current leads the aim is to minimize the heat load introduced when transmitting a given current into the cryostat. The heat load in this case has a second source in addition to pure conduction, which is the ohmic loss in the lead. The goal of the design is therefore to minimize both heat conduction and electrical resistance.

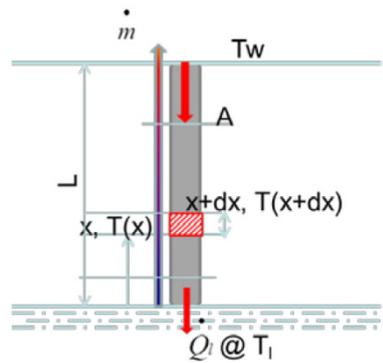

**Fig. 25:** Vapour-cooled electrical conductor

The thermal balance of an element of the conductor $\mathrm{d}x$ (Fig. 25) can be written as:

$$\frac{\mathrm{d}}{\mathrm{d}x}\left( k(T) \cdot A \cdot \frac{\mathrm{d}T}{\mathrm{d}x} \right) - f \cdot \overset{\bullet}{m} \cdot Cp(T) \cdot \frac{\mathrm{d}T}{\mathrm{d}x} + \rho(T) \cdot \frac{I^2}{A} = 0 \quad . \tag{39}$$

where $I$ is the current in the conductor, $\rho$ is its electrical resistivity, and where $f$ is a cooling efficiency factor that can vary between 0 and 1, where 0 corresponds to no cooling and 1 to perfect heat exchange between gas and wall.

For current lead materials that follow the Wiedemann–Franz law (most metals and alloys), thermal conductivity and electrical resistivity are bound by the relation:

$$\rho(T) \cdot k(T) = L_0 \cdot T \ , \qquad (40)$$

which can be inserted into Eq. (39), and solved numerically to calculate the heat load per unit current for varying values of $f$ as plotted in Fig. 26. Perfectly cooled leads can reach the minimum value of ~ 1 W kA$^{-1}$. It should be noted that this result is independent of the conductive material of the conductor. An even better performance can be achieved by using materials that do not follow the Wiedemann–Franz law, for example High-Temperature Superconductors (HTS), having close to zero resistivity and which behave as relatively bad thermal conductors up to high temperatures.

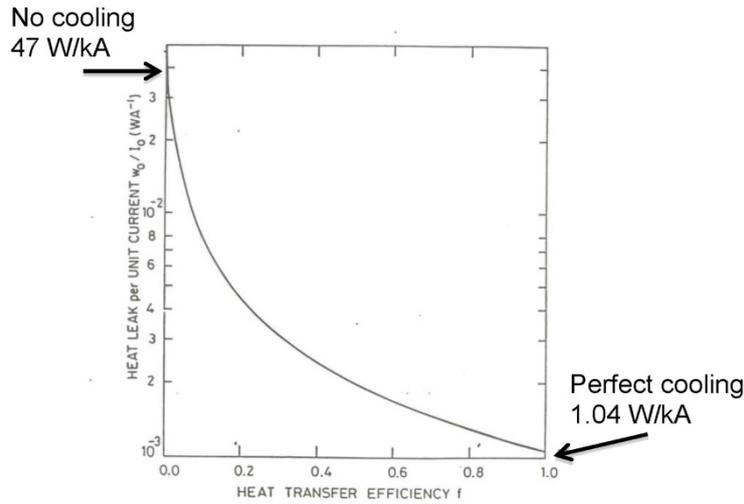

**Fig. 26:** Heat load per unit current in vapour-cooled leads

Vapour cooling is also employed for RF power couplers. An illustrative example is the SPL cryo-module, under development at CERN (Fig. 27). The external conductor of the coaxial RF coupler is made out of a double-walled stainless steel tube with a copper-plated internal diameter of 100 mm; a forced circulation of helium vapour from 5–293 K inside the double-walled envelope, regulated in mass flow and output temperature, allows intercepting conduction of heat as well as resistive RF power deposited in the copper-plated wall.

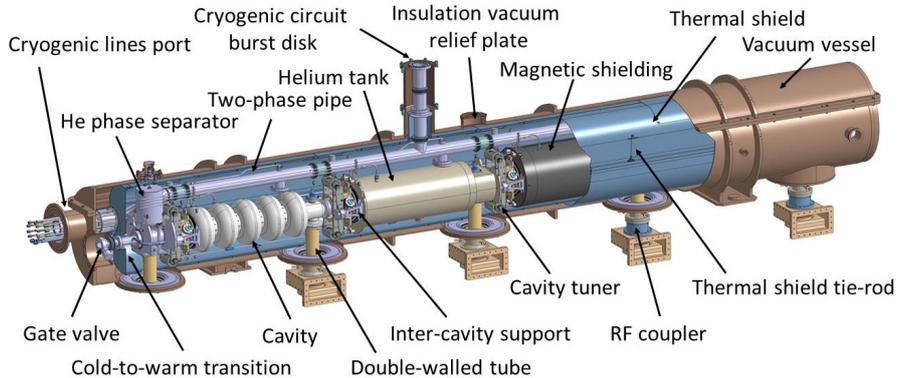

**Fig. 27:** General assembly view of the SPL short cryo-module with vapour-cooled RF couplers

A comparison between heat intercept and vapour-cooled solutions is given in Table 10, where one can appreciate the efficiency of those solutions adopted (cases E and F).

**Table 10:** Cooling efficiency comparison of heat intercepts and vapour cooled solutions for the SPL RF coupler

| Case | | $Q$ at 2 K (W) | Wel (W) | $Q$ at 8 K (W) | Wel (W) | $Q$ at 80 K (W) | Wel (W) | Vapour rate (W) | $Q$ equivalent at 4.5 K (1 g s$^{-1}$ = W) | Wel (W) | Total Wel (W) |
|---|---|---|---|---|---|---|---|---|---|---|---|
| A | No intercept | 11.629 | 11512.71 | | | | | | | | 11,513 |
| B | One optimized and perfect intercepts at 80 K | 1.816 | 1,797.84 | | | 39.513 | 632.208 | | | | 2,4301 |
| C | Two optimized and perfect intercepts at 80 K and 8 K | 0.129 | 127.71 | 2.64 | 580.8 | 26.816 | 429.056 | | | | 1,138 |
| D | 4.5 K self-sustained vapour cooling | 0.031 | 30.69 | | | | | 0.019 | 1.9 | 407 | 438 |
| E | Real case, He vapour cooling 4.5–300 K | 0.1 | 99 | | | | | 0.04 | 4 | 880 | 1,039 |
| F | Real case, He vapour cooling, 4.5–300 K, RF power on | 0.5 | 495 | | | | | 0.04 | 4 | 880 | 1,435 |
| G | Real case, no He vapour cooling, RF power on | 22 | 21,780 | | | | | 0 | 0 | 0 | 21,780 |

# 6    Mechanical and construction aspects

## 6.1    Choice of materials for cryostats

A cryostat is composed, as a minimum, of the following main components:

- Vacuum vessel;
- Cryogenic vessels containing superconductor devices and/or containing cryogens;
- Thermal shield;
- Supporting systems.

Vacuum and cryogenic vessels are sheet-metal constructions, with thicknesses typically in the 1–15 mm range. Materials preferentially range from low-carbon construction steels to stainless steels, though aluminium is also sometimes employed. Vacuum vessel materials operate at room temperature but must be qualified to withstand sudden cool-down to about −70°C in case of accidental rupture of the insulation vacuum. Charpy qualification tests at low temperature have to demonstrate adequate energy absorption.

Cryogenic vessels are generally made of austenitic stainless steels. The austenitic structure does not undergo any ductile-to-brittle transition at cryogenic temperature. The 304-type (18Cr8Ni on a Fe basis) is the most common grade in use. Its low-carbon steel version 304L (C ≤ 0.030%), is preferred because of its enhanced corrosion resistance and ductility in welded structures. Austenite is non-magnetic, but for special applications where a fully austenitic and stable structure is sought even at cryogenic temperatures (high field quality in superconducting magnets, for instance), 316-type steels and, in particular, 316LN should be used. The 316LN type is also commonly used in bellows convolutions owing to its formability and ductility.

Thermal shields must be in high thermal conductivity material. Copper and aluminium are the materials of choice, preferentially as alloys for their better mechanical properties and manufacturability while still preserving good thermal conductivity. Pure coppers are preferred only for demandingly high thermal conductivity where temperature homogeneity matters. Aluminium alloys are by far the most used materials in thermal shields for accelerator cryostats. Series 6000 (Al-Mg-Si), combines good manufacturability (can be extruded) and weldability, with excellent thermal conductivity at 80 K (close to Al1100, pure aluminium).

Materials for supporting systems usually depend on their design. Column-type supports are extensively used in accelerator cryostats, and are the choice for the LHC cryostats. Fibre-reinforced plastic materials are employed because they are good thermal insulators, but their types and material composition, as well as manufacturing processes, cover a wide range of mechanical and thermal properties. Epoxy-based composites, in particular cryogenic-grade glass-fibre composite materials G10 and G11 are widely employed. Injection-moulded plastics, reinforced with short fibres (Ultem) are also materials that have been used.

The LHC support posts are composed of Resin Transfer Moulding (RTM) monolithic 4 mm-thick tubular columns in GFRE material. This material was chosen for its low thermal conductivity-to-stiffness ratio while remaining a commercially available material to keep reasonable costs. Supporting systems based on suspension rods can be made of stainless steel or titanium alloys.

The choice of the materials and the design of the components are primarily dictated by their requirements, but one should not neglect their manufacturability and the capability of having them produced in an industrial context, in particular for large series production, where cost containment from economy of scale and commercial competition can be substantial.

Whenever possible, international standards and norms should be employed and referred to when specifying materials, manufacturing processes, and quality control. The use of commercially available materials, sometimes at the cost of relaxing excessively stringent initial requirements, can be an advantageous strategy to contain costs and widen the number of potential suppliers to foster competition. For instance for cryogenic fluid envelopes 316 LN stainless steel (particularly indicated for superconducting magnet envelopes for preserving low magnetic susceptibility) can be suitably replaced with 304L (a general-purpose grade) at half the material cost. Contrarily, when special applications call for specific material composition and semi-finished material forms, one should make a careful selection and tighten requirements as compared to standard industrial ones. In vacuum flanges, for instance, which are normally machined out of forged bars or cold-rolled plates, non-metallic inclusions in the base material can be stretched and align in directions along which leaks may develop. For applications with stringent tightness requirements, suitable specifications should limit the allowed inclusion content, and the material checked according to microscopic test methods. Specifying the melting process, such as imposing remelting such as Vacuum Arc Remelting (VAR) or ElectroSlag Remelting (ESR), are ways of ensuring the homogeneity of the material. Specifying that flanges are made through three-dimensional forging is also beneficial to avoid inclusions aligning in preferential directions.

When dealing with vacuum and pressure vessels, including piping and thin-walled flexible components (bellows and flexible hoses), the design engineer should not neglect safety aspects and should refer to the relevant regulations in force in the country of construction and operation of the equipment.

## 6.2 Thin shells under pressure

Circular tubes are the most commonly used geometries for internal pressure. Simple formulas can be used in a preliminary design to calculate the main stresses in *thin shells* (where $t < r/10$, with thickness $t$ and radius $r$).

The *circumferential stress* in the presence of an internal pressure $p$ is given by:

$$\sigma_1 = \frac{p \cdot r}{t} \tag{40}$$

whereas the axial force induced by pressure on the closed extremities introduce an *axial stress* that is half the circumferential stress:

$$\sigma_2 = \frac{p \cdot r}{2 \cdot t} \tag{41}$$

and the stress perpendicular to the wall can be neglected: $\sigma_3 \approx 0$.

Considering the *Tresca failure criterion*, and introducing the admissible stress $\sigma_a$, which is a property of the material, we can write:

$$\sigma_a \geq |\sigma_1 - \sigma_3| = \frac{p \cdot r}{t}, \tag{42}$$

which can be used to calculate, for a given material, pressure, and radius, the minimum thickness $t$. This formula holds for both internal and external pressure. Taking for example a vessel of a diameter of 1 m, and with $\sigma_a = 150$ MPa (value for 304L, with a safety factor of 1.5 on $Rp_{1.0}$), for a pressure of 1 bar (0.1 MPa), we can calculate $t > 0.3$ mm.

But when subject to external pressure, thin shells must be designed against radial buckling, which is a non-linear phenomenon depending on initial geometrical imperfections and which can lead to a sudden collapsing of the cylinder when a *critical pressure* is exceeded, as illustrated in Fig. 28.

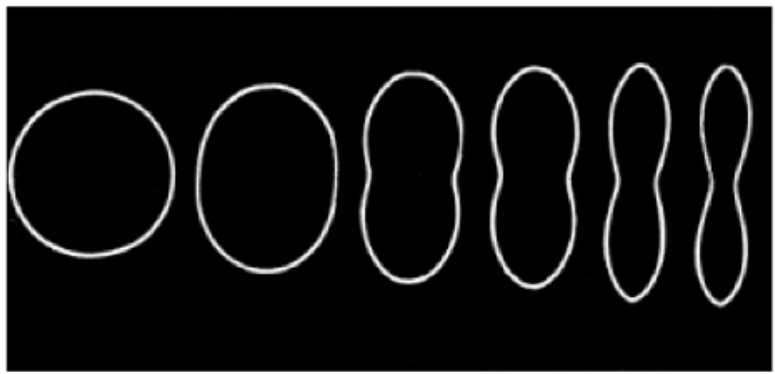

**Fig. 28:** Radial buckling of a thin tube

For a thin tube of infinite length, the critical pressure is given by:

$$p_{cr} = \frac{E}{4(1-\upsilon^2)} \cdot \left(\frac{t}{r}\right)^3 \tag{43}$$

with $E$ and $\upsilon$ the Young's modulus and Poisson ratio of the material, respectively.

As a rule of thumb, thickness and radius should satisfy the ratio $t/r > 0.012$. To cope with geometrical imperfections, a minimum safety factor of 3 should be included, so the design rule becomes:

$$\frac{t}{r} \geq 3.7\% \ . \tag{44}$$

For instance, for the LHC cryostat vacuum vessel, with a radius of about 0.5 m, the thickness should be at least 18.5 mm (in fact it could be reduced to 12 mm by introducing stiffening ribs).

### 6.3 Supporting systems in accelerator cryostats

Supporting systems in accelerator cryostats are designed to support and precisely position superconducting devices. Magnets generally require alignment precisions within few tenths of a millimetre, whereas SRF cavities are less demanding, but still require precisions within fractions of a millimetre. Alignment must be maintained irrespective of thermo-mechanical movements induced by the cool-down and warm-up transients, and drifts throughout the thermal cycles in the lifetime of an accelerator.

Adjusting the position of each entire cryostat by means of moving them on their external supporting system is the easiest way of aligning the superconducting device, provided the cryostat internal supporting system keeps its positioning stability and is reproducible in time.

The LHC supporting system is based on this principle, as are most of the large accelerators (for instance the Hadron Electron Ring Accelerator (HERA), Relativistic Heavy Ion Collider (RHIC), and Spallation Neutron Source (SNS).

In some machines, the internal supporting systems can be adjusted from outside the cryostat; in this case the positioning stability requirement is less stringent but a position monitoring system of the superconducting device should be included if adjustment in cold operation is desired. For instance, the HIE Isolde cryomodules, under construction at CERN, integrate a motorized adjustment of the internal structure supporting the cavities and solenoids, and a CCD-based optical camera monitoring their internal position through view ports (Fig. 29).

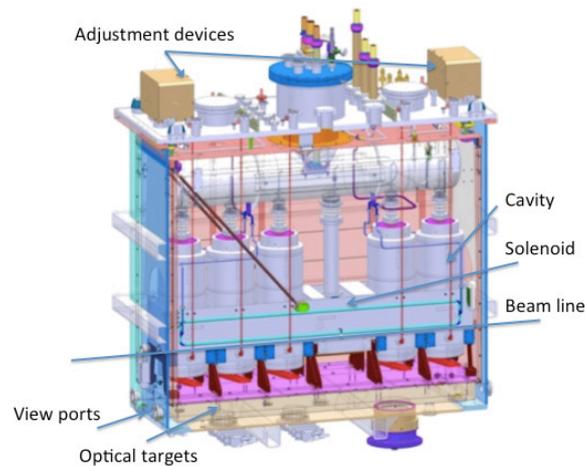

**Fig. 29:** HIE ISOLDE cryomodule with optical monitoring and external adjustment of cavities/solenoid

Various supporting solutions can be employed; a non-exhaustive overview is given below. The LHC is based on column-type compression posts, in GFRE material; RHIC makes use of a double column-type compression support in Ultem™, a polyetherimide thermoplastic material (Fig. 30).

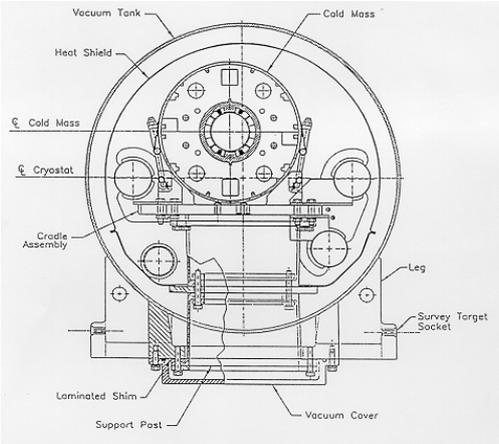

**Fig. 30:** RHIC dipole on column-type supports

A supporting system making use of vertical suspension and transversal adjustment tie rods was adopted for the HERA dipoles (Fig. 31). The position of the magnet could be adjusted from outside the cryostat, but only when warm.

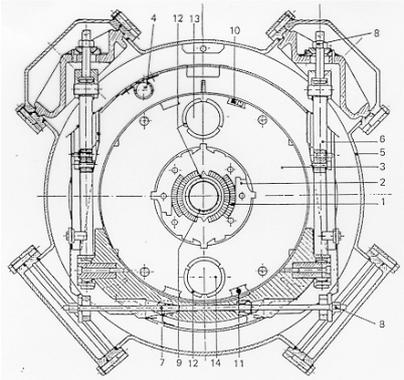

**Fig. 31:** HERA dipole, with suspension tie rods

The X-Ray Free Electron Laser (XFEL) cryomodule also makes use of column-type posts, but in a suspended configuration (Fig. 32).

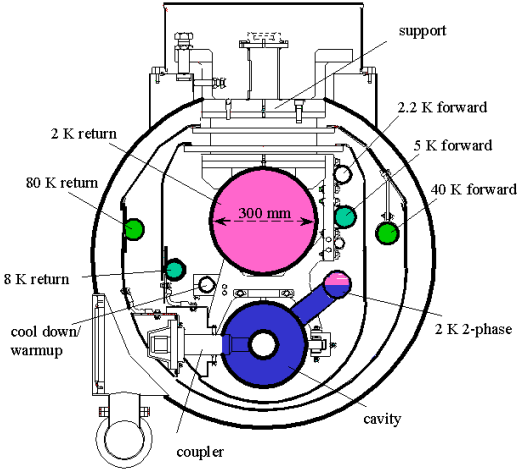

**Fig. 32:** XFEL cryomodule, with suspended column-type supports

The SNS cryomodules, based on the Continuous Electron Beam Accelerator Facility (CEBAF) design, makes use of a frame to which cavities are suspended and adjusted vie tie rods, and which is suspended inside the vacuum vessel (Fig. 33).

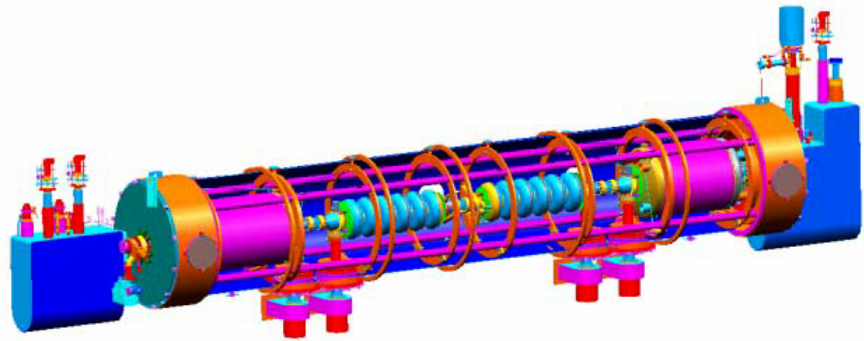

**Fig. 33:** SNS cryomodule with internal frame and tie rods suspension of cavities

Finally, the Superconducting Proton Linac (SPL) cryomodule, under development at CERN, proposes the use of the double-walled tube of the external conductor of the RF power coupler as the main mechanical support for the cavities, with the addition of a secondary supporting element between adjacent cavities (Fig. 34).

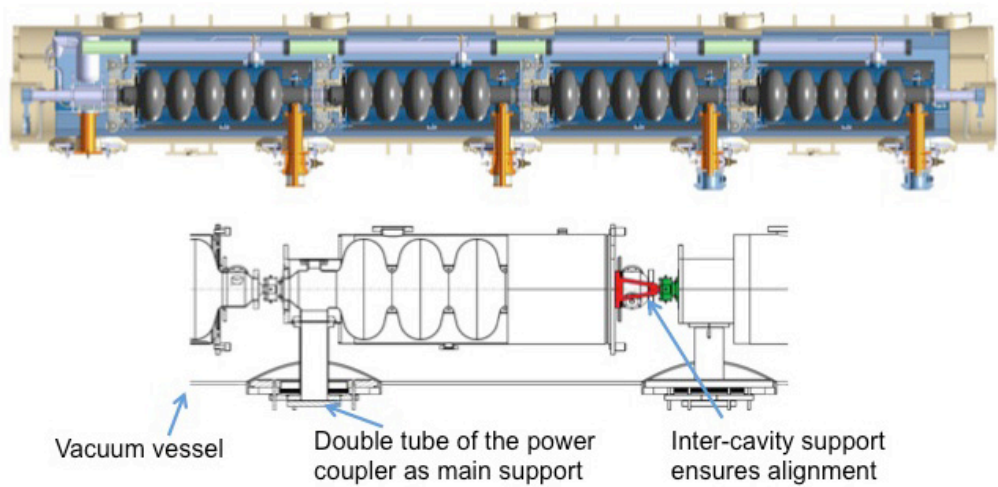

**Fig. 34:** SPL cryomodule. RF coupler's double-walled tube as main cavity support

### 6.4 Welded and brazed assemblies

In the fabrication of cryostat envelopes and components, welding of steel and aluminium is a common technology, which achieves two basic and distinct functions: providing structural reinforcement in an assembly, or ensuring leak-tight joining. The latter creates a non-dismountable assembly and is chosen for permanent assemblies. Nevertheless, welding can be a cost-effective solution, when compared to flanged assemblies, for joints that only need occasional dismounting. For instance, the LHC magnet interconnections, where the cryogenic piping contains the electrical connections, are all welded, despite the occasional need for replacing magnets.

Leak-tight welding requires special precautions in preparation and execution, and must follow very strict protocols and Quality Control (QC) checks. The preferred and most common technology is Tungsten Inert Gas (TIG), though Metal Inert Gas Welding (MIG), is also in use (Fig. 35).

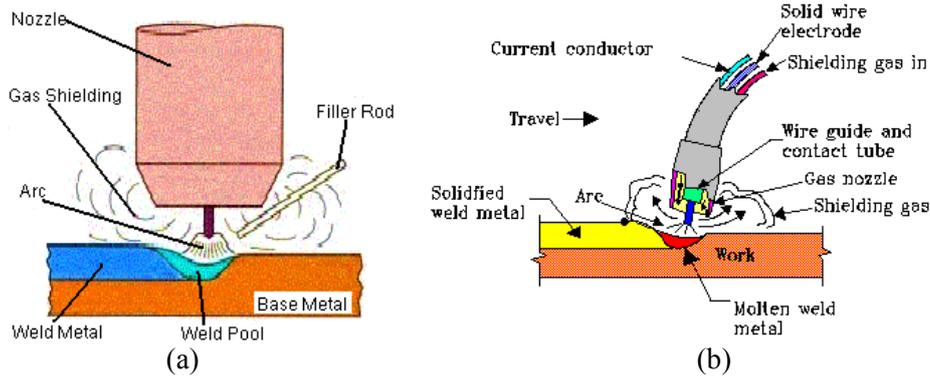

(a)                                       (b)

**Fig. 35:** Welding technologies. (a) TIG; (b) MIG

When TIG is employed in leak-tight welding of thin-walled structures, filler material is often not necessary. In designing weld joints, one should follow prescribed geometries (Fig. 36 shows some of these rules). Leak-tight welds should always be on the vacuum side of the component. Structural reinforcing welds in the presence of a leak-tight weld shall always be made on the external side and discontinuously, not trapping closed volumes between the two welds. One of the keys to successful welding is an adequate protection by backing gas (argon or helium). Adequate protection is ensured by the TIG torch on the welding side, but proper protection should also be provided on the back side. Design of welded seams must be carefully chosen to avoid sources of impurities and defects, especially for UHV applications. Non-destructive inspections like X-rays are often necessary on structural welds, which should be designed to be accessible for inspection.

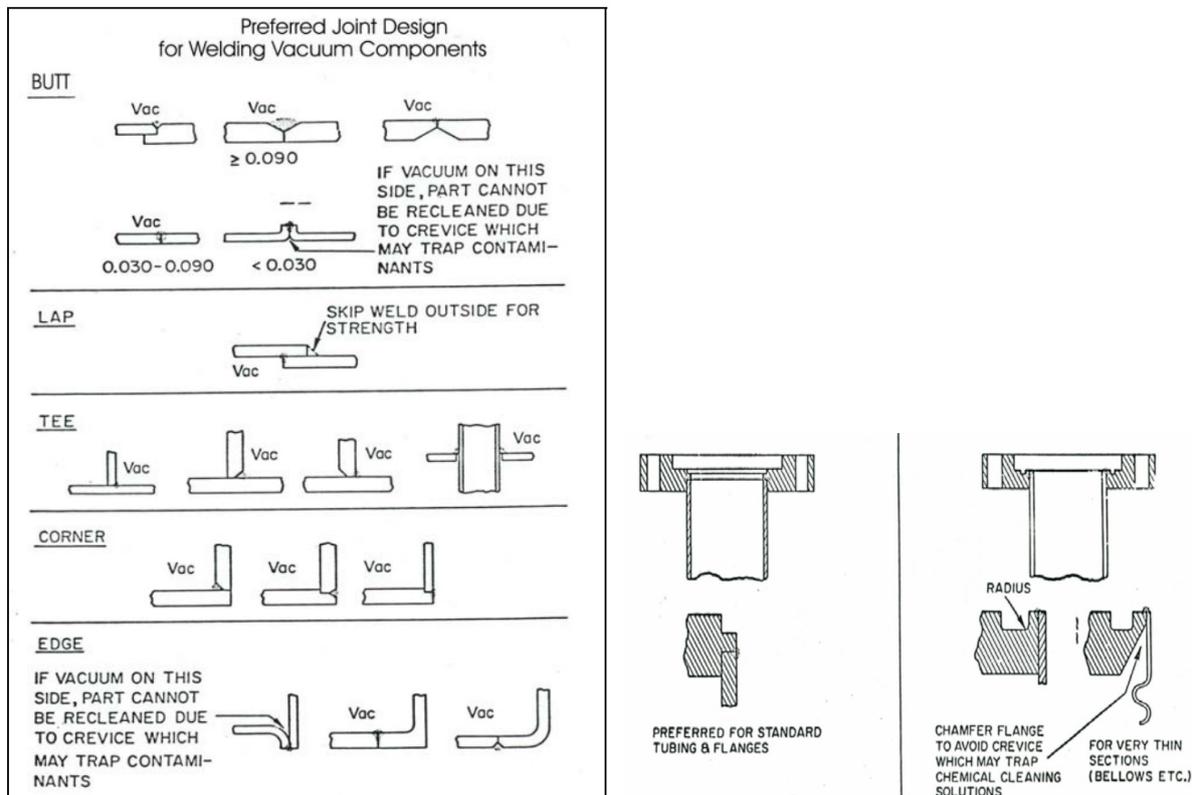

**Fig. 36:** Welding geometries for vacuum leak-tight welds

Welds are normalized and should follow welding procedure qualification, especially if the welds have a structural function on pressure vessels (Table 11 reports the European Norms is force). In this case the welds' design should conform to the pressure vessel codes in use; welders are also required to be qualified for the execution of specific weld geometries. Qualification of welders is carried out by accredited qualification bodies.

**Table 11:** European Norms relevant to steel and aluminium welds and welders

| | Steel | Aluminium |
|---|---|---|
| Welding procedure approval | EN ISO 15614-1:2004 Specification and qualification of welding procedures for metallic materials—Welding procedure test—Arc and gas welding of steels and arc welding of nickel and nickel alloys | EN ISO 15614-1:2005 Specification and qualification of welding procedures for metallic materials—Welding procedure test—Arc and gas welding of steels and arc welding of aluminium and alloys |
| Qualification of welders | EN 287-1:2004 Qualification test of welders—Fusion welding—Steels | EN ISO 9606-2:2004 Qualification test of welders—Fusion welding—Aluminium and aluminium alloys |
| Qualification of welding operators | EN 1418:1998 Welding personnel—Approval testing of welding operators for fusion welding and resistance weld setters for fully mechanized and automatic welding of metallic materials | |

Brazing is an assembly alternative for materials that cannot be welded, or for dissimilar materials (copper on steel, for instance). It consists of heating a filler material just above its melting point so that it can flow by capillary action between the two materials to be joined. The use of flux agents is essential to avoid oxides from forming and degrading the quality of adhesion. A proper cleaning after brazing is essential to eliminate flux residues, which can provoke corrosion and possibly cause leaks in the base materials.

Brazing can be carried out in furnaces under controlled atmosphere, but this limits the size of the parts to be assembled and is costly.

A list of available European Norms covering brazing is given below:

- EN 13134:2000 Brazing—Procedure approval;
- EN 13133:2000 Brazing—Brazer approval;
- EN 12797:2000 Brazing—Destructive tests of brazed joints;
- EN 12799:2000 Brazing—Non-destructive examination of brazed joints;
- EN ISO 18279:2003 Brazing—Imperfections in brazed joints.

### 6.5 Pressure vessels and construction standards

Since 2002, the Pressure European Directive 97/23/EC (PED) has become an obligatory legislation throughout the EU. It applies to all equipment with an internal *maximum allowable pressure* ≥0.5 bar, in which case the equipment is considered to be a *pressure vessel*. Vessels must be designed, fabricated, and tested according to the essential requirements defined in the PED (design, safety accessories, materials, manufacturing, testing, etc.), and must be subject to a qualification pressure test under 1.43 times the maximum allowable pressure.

The category of the pressure vessel (from I to IV) depends on the stored energy, expressed as pressure × volume in bar.l, according to the diagram in Fig. 37. The category establishes the conformity assessment procedure to be adopted. Vessels not falling under the categories are to be designed in accordance with sound engineering practice. From category I and above, CE marking becomes obligatory and the level of conformity assessment increasingly stringent (Table 12).

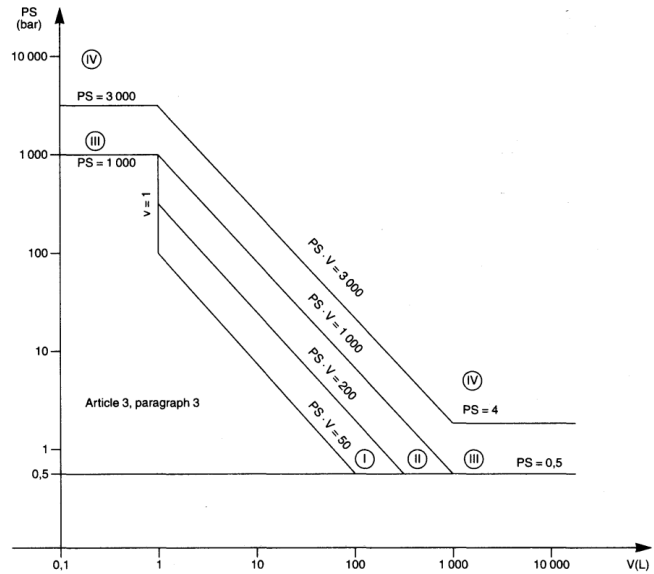

**Fig. 37:** Category of pressure vessels according to stored energy expressed in bar.l

**Table 12:** Category of pressure vessels and associated conformity assessments

| Category | Conf. assessment module | Comment |
|---|---|---|
| SEP | None | The equipment must be designed and manufactured in accordance with sound engineering practice. No CE marking and no involvement of notified body. |
| I | A | CE marking with no notified body involvement, self-certifying. |
| II | A1 | The notified body will perform unexpected visits and monitor final assessment. |
| III | B1 + F | The notified body is required to approve the design, examine and test the vessel. |
| IV | G | Even further involvement of the notified body. |

When designing pressure equipment, the engineer should make use of, as far as applicable, standards and norms, as this approach gives presumption of conformity with the PED.

A list of useful norms for cryostat design and fabrication is given below:

- EN 13458-1:2002 Cryogenic vessels—Static vacuum insulated vessels—Part 1: Fundamental requirements;

- EN 13458-2:2002 Cryogenic vessels—Static vacuum insulated vessels—Part 2: Design, fabrication, inspection and testing + EN 13458-2:2002/AC:2006;

- EN 13458-3:2003 Cryogenic vessels—Static vacuum insulated vessels—Part 3: Operational requirements + EN 13458-3:2003/A1:2005;

- EN 13445-1:2009 Unfired pressure vessels—Part 1: General;

- EN 13445-2:2009 Unfired pressure vessels—Part 2: Materials;

- EN 13445-3:2009 Unfired pressure vessels – Part 3: Design;

- EN 13445-4:2009 Unfired pressure vessels – Part 4: Fabrication;

- EN 13445-5:2009 Unfired pressure vessels – Part 5: Inspection and testing;

- EN 13445-8:2009 Unfired pressure vessels – Part 8: Additional requirements for pressure vessels of aluminium and aluminium alloys.

Other design codes such as the French CODAP or the American ASME Boiler and Pressure Vessel Code can be used, but proof of conformity is at the charge of the manufacturer.

Furthermore, useful European Norms for semi-finished products are listed in Table 13.

**Table 13:** List of European Norms for semi-finished products

| Product | European Norm |
|---------|---------------|
| Plate and sheets | EN 10028-1:2007+A1:2009 Flat products made of steels for pressure purposes—Part 1: General requirements |
| | EN 10028-3:2009 Flat products made of steels for pressure purposes—Part 3: Weldable fine grain steels, normalized |
| | EN 10028-7:2007 Flat products made of steels for pressure purposes—Part 7: Stainless steels |
| Tubes | EN 10216-5:2004 Seamless steel tubes for pressure purposes—Technical delivery conditions—Part 5: Stainless steel tubes |
| | EN 10217-7:2005 Welded steel tubes for pressure purposes—Technical delivery conditions—Part 7: Stainless steel tubes |
| Forged blanks | EN 10222-1:1998 Steel forgings for pressure purposes—Part 1: General requirements for open die forgings |
| | EN 10222-5:1999 Steel forgings for pressure purposes—Part 5: Martensitic, austenitic and austenitic-ferritic stainless steels |
| Castings | EN 10213:2007 Steel castings for pressure purposes |
| Pipe fittings | EN 10253-4 Butt-welding pipe fittings—Part 4: Wrought austenitic and austenitic-ferritic (duplex) stainless steels with specific inspection requirement |
| Bars | EN 10272:2007 Stainless steel bars for pressure purposes |
| Aluminium | EN 12392:2000 Aluminium and aluminium alloys –Wrought products—Special requirements for products intended for the production of pressure equipment (choose materials included in the list given in EN 13445-8 section 5.6) |

# 7 Pressure relief protection systems

Cryostats contain large cold surfaces, cryogenic fluids, and sometimes large stored energy (e.g. energized magnets), with the potential risk of sudden liberation of energy through thermodynamic transformations of the fluids, which can be uncontrolled and lead to a dangerous increase of pressure inside the cryostat envelopes.

The consequence, in the case of a rupture of the envelopes, may be serious for personnel (injuries from deflagration, burns, and oxygen deficiency hazard) as well as for the equipment.

Performing a thorough risk analysis is essential to identify and understand risk hazards that may cause a pressure increase and in order to assess consequences, define mitigation actions, and design adequate safety relief devices to limit pressure accordingly.

A non-exhaustive list of potential sources of pressure increase is:

- compressors connected to cryogenic lines;

- connection to higher pressure source (e.g. HP bottles);

- heating of 'trapped' volumes (typically in a circuit between valves) during warm-ups;

- helium leak to insulation vacuum, with consequent increased conduction/convection heat loads to cryogenic liquid vessels;

- cryo-condensed air leaks on cold surfaces and consequent pressure increase and increased conduct/convection heat loads during warm-ups;

- heating/vaporization of cryogens from sudden release of stored energy in superconductor device (e.g. quench or arcing in a superconductor magnet circuit);

- uncontrolled air/nitrogen venting of insulation vacuum with sudden condensation on cold surfaces;

- uncontrolled release of cryogenic fluid to higher temperature surfaces (thermal shield and vacuum vessel), and consequent pressure increase and increased conduction/convection heat loads to cold surfaces.

The last three of these causes are generally those that define the sizing conditions for the overpressure relief devices in cryostats for superconducting devices, and are developed in the following section. The applicable European Norm is EN 13648 Safety devices for protection against excessive pressure.

## 7.1 Design of pressure relief devices

The design of pressure relief devices should be made for each cryostat envelope that is to be protected; it includes the vacuum vessel and all contained cryogenic reservoirs and circuits.

The design approach can be subdivided as follows:

a) Risk analysis and mitigation:

- Make a thorough risk analysis and *evaluate risk hazards*;

- *Identify mitigation measures* (e.g. protection of exposed bellows and flanged connections);

- *Identify severity of consequences* and *appreciate probability* of the event;

- Define the *Maximum Credible Incident(s) (MCI)* and design the safety relief system accordingly. The safety relief system must be designed to keep pressure rise within the limits of the Maximum Allowable Working Pressure (MAWP).

b) Design steps for MCI:

- Estimate the heat exchange and its *conversion to mass flow rates* to be discharged;

- *Check the sizing of piping* (generally designed for normal operation) to the relief device and increase if necessary;

- *Choose the type of safety device* (burst disks, valves, plates) and size the safety device (DN and set pressure). Make use of safety device manufacturers' formulas and charts;

- *Size recovery piping downstream of safety device* and check venting needs in the building where the release occurs (Oxygen Deficiency Hazard issue).

### 7.1.1 Design of pressure relief devices for cryogenic fluid vessels

Cryogenic fluid vessels are, in general, pressure vessels, where the MAWP can be in the range of a few bar (for SRF cavity vessels) up to a few tens of bar (for magnet vessels).

Let's consider that the hazard leading to the MCI is a breach in the insulation vacuum, leading to a fast venting with air of the vacuum vessel volume, with sudden condensation heat transfer on the cold surfaces, and boil-off of the cryogen with an increase of pressure (Fig. 38).

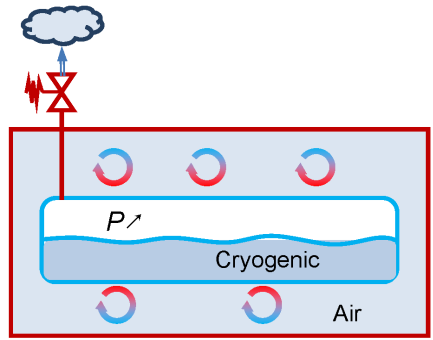

**Fig. 38:** Sudden air venting of the insulation vacuum volume, pressure increase in cryogenic vessel, relieving through its safety device.

The heat flux exchange on the cold surface has been measured experimentally, and reference values for different insulation configurations can be found in literature. The following conservative values can be used:

- 0.6 W cm$^{-2}$ for a vessel protected by ten layers of MLI;
- 4 W cm$^{-2}$ for a bare surface.

The safety relief device should be designed to be able to relieve a mass flow equivalent to the highest heat load.

The mass flow, $Q_m$ can be calculated by following the approach described in EN13468-3.4.

Formulas are given to distinguish between cases in which pressure remains below or exceeds the critical pressure ($p = 2.23$ bar for helium).

The same norm provides, under conservative assumptions, formulas for calculating the required flow relief cross-section, depending on whether the flow is critical or sub-critical.

This is the minimum cross-section area of the safety relief device to be chosen. Figure 39 illustrates a few examples of commercially available pressure relief devices: a burst disk, calibrated to break at a given pressure difference, and a reversible spring-loaded valve.

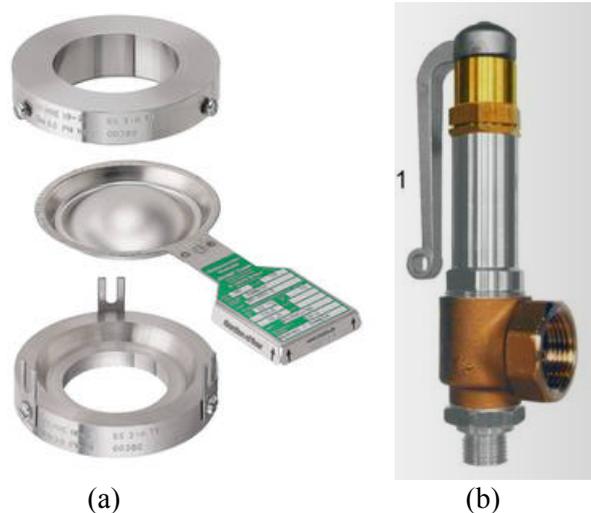

(a)                                      (b)

**Fig. 39:** Pressure relief devices. (a) burst disk; (b) spring-loaded valve

### *7.1.2 Design of pressure relief devices for vacuum vessels*

According to the Pressure European Directive 97/23/EC, in order not to be considered a pressure vessel, vacuum vessels should be designed to keep the internal MAWP within 1.5 bar absolute (i.e. $\Delta p < 0.5$ with respect to atmospheric pressure) in all working conditions, including an accidental event. The safety relief devices are therefore to be designed to keep within these values.

Often the MCI corresponds to the rupture of a cryogenic circuit: the cryogenic liquid floods the vacuum vessel, vaporizes and expands in contact with the warm walls, the internal pressure increases until the safety relief device opens, relieving the fluid to atmospheric pressure (Fig. 40). The vacuum vessel safety device is to be designed to relieve a mass flow equal to the highest flow from the cryogenic vessel, but at a warmer temperature, while keeping the vacuum vessel pressure within 1.5 bar absolute.

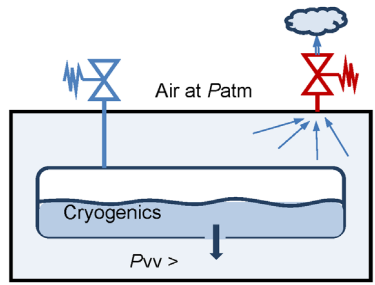

**Fig. 40:** Breach in the cryogenic fluid vessel, expansion in the vacuum vessel and relief through its safety device.

Calculating the maximum mass flow from the cryogenic fluid vessel to the vacuum vessel consists of first estimating the size of the breach in the cryogenic circuit (for instance rupture of a bellows), then calculating the mass flow through the orifice, depending on the flow conditions and pressure inside the cryogenic vessel.

Conservatively, the cross-section of the vacuum vessel safety device should be able to discharge the same mass flow, but at a warmer temperature. This cross-section is highly dependent on the relief temperature, which is difficult to estimate. In the first instance one should consider the most conservative assumption of discharge at room temperature. If the sizing results are excessively large (often the case), one should review this assumption by taking into account more realistic (but still conservative) assumptions limiting the warm-up of the fluid along the evacuation path, for instance assuming that the thermal shield protects the fluid from heat-up too much. Once the minimum relief cross-section is calculated, this should be used to select the safety relief device. Figure 41 illustrates the DN200 spring-loaded safety relief plate mounted on each LHC dipole cryostat.

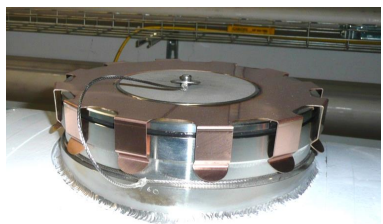

**Fig. 41:** LHC dipole cryostat. DN200 spring-loaded safety relief plate

## 8    Summary

After more than one century of developments and applications since the first laboratory cryogenic reservoirs of J. Dewar, cryostats are nowadays industrially produced. From a first intuitive understanding of heat exchange, we now have a solid understanding of the phenomena involved. An overview of the basic thermal and mechanical design aspects has been presented, trying to provide simple calculation tools for preliminary design.

If cryostats are built in industrial series for commercial applications, accelerator cryostats remain a niche in the field, and are designed for very specific applications in superconductivity, most of the time requiring state-of-the-art enabling technologies like cryogenics and vacuum. Some aspects of these disciplines were treated. Mechanical and construction issues have been mentioned, as well as standards and norms, which apply when cryostats are fabricated in an industrial context. The LHC cryostat has been extensively used as a case study throughout this paper. For further developments, the reader is invited to refer to the bibliography presented hereafter.

## Acknowledgements

The work presented in this course is essentially the result of contributions from a number of colleagues and the work done during the design and construction of the LHC. I wish to acknowledge in particular for the material provided and for their contributions in preparing this course: R. Bonomi, P. Cruikshank, Ph. Lebrun, Y. Leclercq, A. Poncet, D. Ramos, and G. Vandoni.

## Appendix A. Case study: Thermal radiation design of the Large Hadron Collider cryostats

Let's consider the design of a tubular cryostat as in the case of the LHC. The aim is to analyse the effects, in a step-by-step sequence, of insulation vacuum, thermal shielding, and MLI on heat loads.

Consider the following data:

- Magnet cold mass: 316 L (stainless steel), diameter 0.6 m, $T = 2$ K;
- Vacuum vessel: low-carbon construction steel, diameter 1 m, $T = 293$ K;
- Heat loads budgets: $\sim 0.2$ W m$^{-1}$ at 2 K; $\sim 5$ W m$^{-1}$ at 80 K;

We will consider both the cases with a good insulation vacuum ($P = 10^{-3}$ Pa ($10^{-5}$ mbar)), and a degraded insulation vacuum ($P = 10^{-1}$ Pa ($10^{-3}$ mbar)).

All calculations in the following are made for a cryostat unit length of 1 m.

### a) Bare cold mass, good insulation vacuum

Considering:

- Emissivity cold mass: $\varepsilon_{CM} = 0.12$
- Emissivity vacuum vessel: $\varepsilon_{VV} = 0.2$

From Eq. (27):

$$Q = \frac{\sigma A_{CM}(T_{VV}^4 - T_{CM}^4)}{\dfrac{1}{\varepsilon_{CM}} + \dfrac{A_{CM}}{A_{VV}}\left(\dfrac{1}{\varepsilon_{VV}} - 1\right)} \tag{A.1}$$

We calculate the heat load to the cold mass:

$$HL_{CM} = 63 \text{ W} \tag{A.2}$$

which is far too high with respect to the budget.

### b) One aluminium foil around cold mass, good insulation vacuum

With:

- Emissivity of Al foil (at 2 K): $\varepsilon_{CM} = 0.06$

We now calculate:

$$HL_{CM} = 40 \text{ W} \tag{A.3}$$

which remains too high.

### c) 30 MLI layers around cold mass, good insulation vacuum

Considering a heat load from 290 K with 30 MLI layers (calculated with the MLI formula): 1.2 W m$^{-2}$ we now calculate:

$$HL_{CM} = 2.3 \text{ W} \tag{A.4}$$

Despite the large gain, we are still one order of magnitude off budget.

### d) Bare cold mass, actively cooled thermal shield, good insulation vacuum

Let us now introduce an intermediate thermal shield in aluminium, actively cooled to 80 K, but still consider a bare cold mass.

- Thermal shield diameter: 0.8 m
- Thermal shield $T = 80$ K
- Emissivity of Al (at 80 K): $\varepsilon_{TS} = 0.1$

Using the equation for radiation, we calculate the heat load to the cold mass and a second heat load to the thermal shield $HL_{TS}$:

$$HL_{CM} = 0.26 \text{ W} \rightarrow \text{close to budget} \tag{A.5}$$

$$HL_{TS} = 79 \text{ W} \rightarrow \text{above budget (5 W m}^{-1}) \tag{A.6}$$

### e) Adding 30 MLI layers around the thermal shield, good insulation vacuum

Considering a heat load from 290 K with 30 MLI layers (calculated with MLI formula): 1.2 W m$^{-2}$:

$$HL_{CM} = 0.26 \text{ W} \rightarrow \text{close to budget} \tag{A.7}$$

$$HL_{TS} = 3.0 \text{ W} \rightarrow \text{within budget} \tag{A.8}$$

### f) Adding one Al foil around the cold mass, good insulation vacuum

With:

- Emissivity of Al foil (at 2 K): $\varepsilon_{CM} = 0.06$

We now calculate:

$$HL_{CM} = 0.18 \text{ W} \rightarrow \text{within budget} \tag{A.9}$$

$$HL_{TS} = 3.0 \text{ W} \rightarrow \text{within budget} \tag{A.10}$$

This configuration satisfies the heat load budgets, provided the insulation vacuum is good.

### g) Adding one Al foil around the cold mass, degraded insulation vacuum

Let us now consider case f) but in a case where, due to a helium leak, the insulation vacuum is degraded. By making use of Eqs. (18) and (19), we can calculate the residual conduction contribution to heat loads.

In a first instance let us consider $P = 10^{-3}$ Pa ($10^{-5}$ mbar), which is still a relatively good vacuum. In this case:

$$Q_{res} = 0.15 \text{ W} \tag{A.11}$$

And the total load is:

$$Q = Q_{rad} + Q_{res} = 0.18 + 0.15 = 0.33 \text{ W} \tag{A.12}$$

Though exceeding the budget the contribution from residual conduction remains limited.

Let us now consider $P = 10^{-1}$ Pa ($10^{-3}$ mbar), which is a bad vacuum. In this case:

$$Q_{res} = 15 \text{ W} \tag{A.13}$$

which is two orders of magnitudes higher and unacceptably high.

### h) Adding 10 MLI layers around the cold mass, good and degraded insulation vacuum

By adding 10 MLI layers on cold mass, and using measured data in good vacuum ($<10^{-3}$ Pa): 50 mW m$^{-2}$, we calculate:

$$HL_{CM} = 0.09 \text{ W} \rightarrow \text{well within budget} \tag{A.14}$$

$$HL_{TS} = 3.6 \text{ W} \rightarrow \text{within budget.} \tag{A.15}$$

Under degraded vacuum ($\sim 10^{-1}$ Pa), and measured data with MLI: $\sim 2$ W m$^{-2}$, we calculate:

$$HL_{CM} = 3.8 \text{ W} \rightarrow \text{above budget} \tag{A.16}$$

$$HL_{TS} = 3.6 \text{ W} \rightarrow \text{within budget.} \tag{A.17}$$

Even though the budget at 2 K is exceeded, it can be noted that MLI on the cold mass reduces heat loads by a factor of 4 in the case of degraded vacuum. Furthermore, MLI is normally wrapped around cold surfaces in order to reduce, by a factor of about 7, condensation heat fluxes in the case of accidental cryostat venting with air (bare surface: $q \sim 4$ W cm$^{-2}$; 10 layers of MLI: $q \sim 0.6$ W cm$^{-2}$).

Table A.1 summarizes the various configurations calculated.

**Table A.1:** Summary of heat loads for various configurations

| Case | 2 K heat loads (W) | 80 K heat loads (W) |
|---|---|---|
| a) Bare cold mass | 63 | NA |
| b) Cold mass with one Al foil | 40 | NA |
| c) Cold mass with 30 MLI layers | 2.3 | NA |
| d) One thermal shield at 80 K, no MLI | 0.26 | 79 |
| e) 30 MLI layers on thermal shield | 0.26 | 3.0 |
| f) As e) + one Al foil on cold mass | 0.18 | 3.0 |
| g) As f) but degraded vacuum[a] | Up to 15 | >33.0 |
| h) +10 MLI layers on cold mass | 0.09[b] | 3.6 |
| | 3.5[c] | >3.6 |

[a]$10^{-1}$ Pa; [b]in good vacuum; [c]in degraded vacuum.

## Appendix B. Experimental assessment of the thermal performance of the Large Hadron Collider cryostats

The following presents the assessment of the static heat loads to the 1.9 K cold mass and to the 50–65 K thermal shielding, carried out during the first commissioning period. Results are then compared with budgets. A sensibility analysis of parameters like insulation vacuum pressure and cryogenic line temperatures is also discussed.

### Cryogenic cooling loop of the arc cells

The cold masses containing the superconducting magnets are cooled via a separate cryogenic circuit, featuring an internal bayonet heat exchanger operating at low-pressure saturated helium (16 mbar), in which liquid is fed from the cryogenic line *C* and vapours are pumped via the cryogenic line *B* back to the cryogenic plant. This cooling loop extends over a total length of 107 m, covering two quadrupoles and six dipoles, and ends in a helium phase separator for collection of liquid in excess (Fig. B.1). The common pressurized 1.9 K static helium bath of the cold masses, a so-called *cryogenic subsector*, extends over the length of two cooling loops (214 m) and is delimited by plugs.

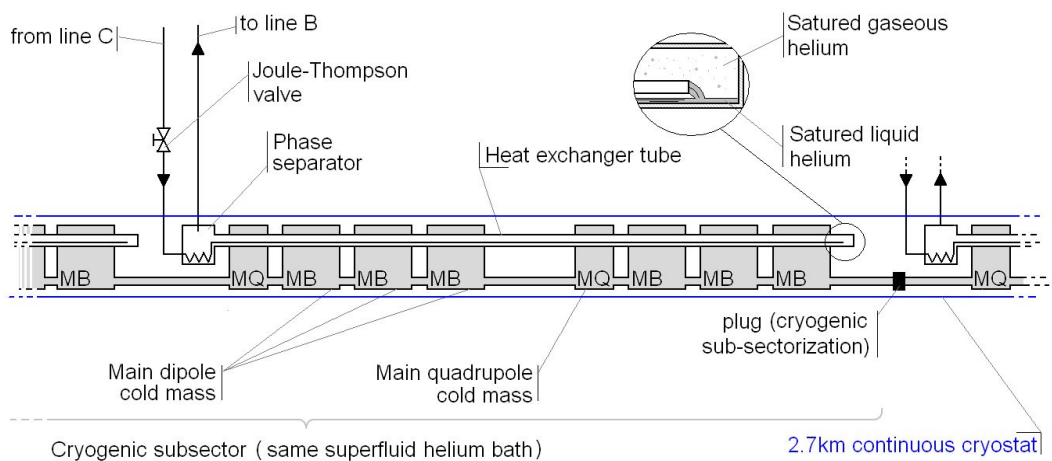

**Fig. B1:** Schematic cryogenic layout of a standard arc cell

A schematic of the heat flow in the LHC cryostat is presented in Fig. B.2.

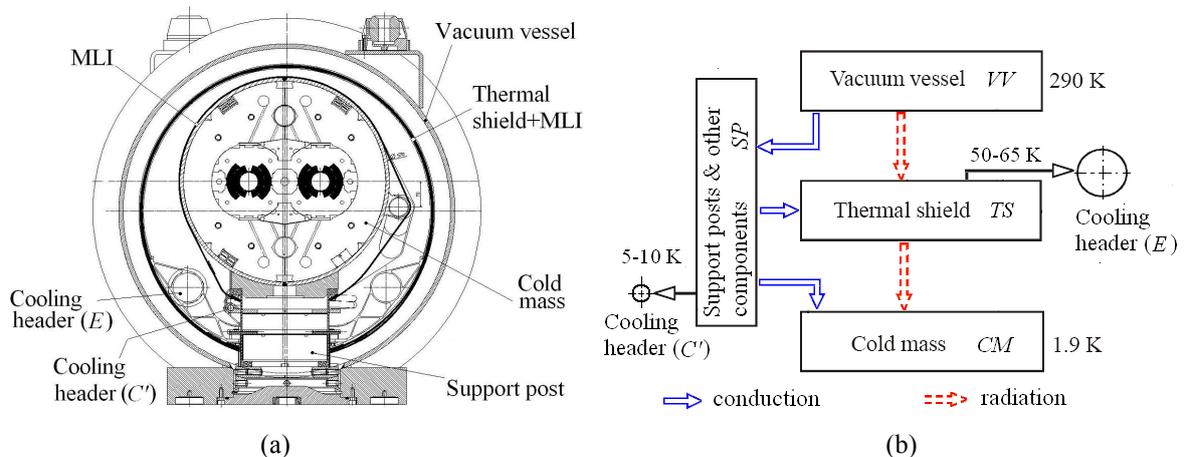

**Fig. B.2:** (a) Cross-section of a dipole cryostat; (b) schematic of heat flows and heat intercept temperatures

**HeII calorimetric measurements of heat load to the cold mass**

The static heat load received by the cold masses in a cryogenic subsector, considered as a closed system, can be estimated by the change in its internal energy over time, during a warm-up, while active cooling via the heat exchanger is stopped. A cryogenic sector can be substantially considered a closed system when the valves to the cryogenic distribution line are closed and leak-tight. This assumption can be checked by calculating the density of helium, on the basis of measured pressure and temperature in the system, and verifying that it remains constant during the isochoric transformation.

Moreover, since the internal energy of the cold masses is derived from point measurements of the temperature and pressure in the system, a uniform temperature is required; this condition is ensured at temperatures below the λ point, taking advantage of the very high heat transport properties in superfluid helium. Calculating the internal energy also requires knowledge of the heat capacity of the cold masses, and in particular related to their helium content, which represents the dominating contribution at cryogenic temperatures. The advantage of the measurement method described below is that it allows estimation of both the helium content and the static heat load by calorimetric measurement.

After stopping the supply of sub-cooled helium to the heat exchanger in one or more subsectors, a natural warm-up (phase *A*) starts. After about 1.5 h, a faster warm-up is forced by injecting a known electrical power (typically 40 W per subsector) into the cold masses via built-in heaters, giving a second trend of about 1 h in the temperature increase (phase *B*). Fig. B.3 illustrates the evolution of temperatures in the cold masses during one of these tests. The constant calculated value of helium density is also plotted.

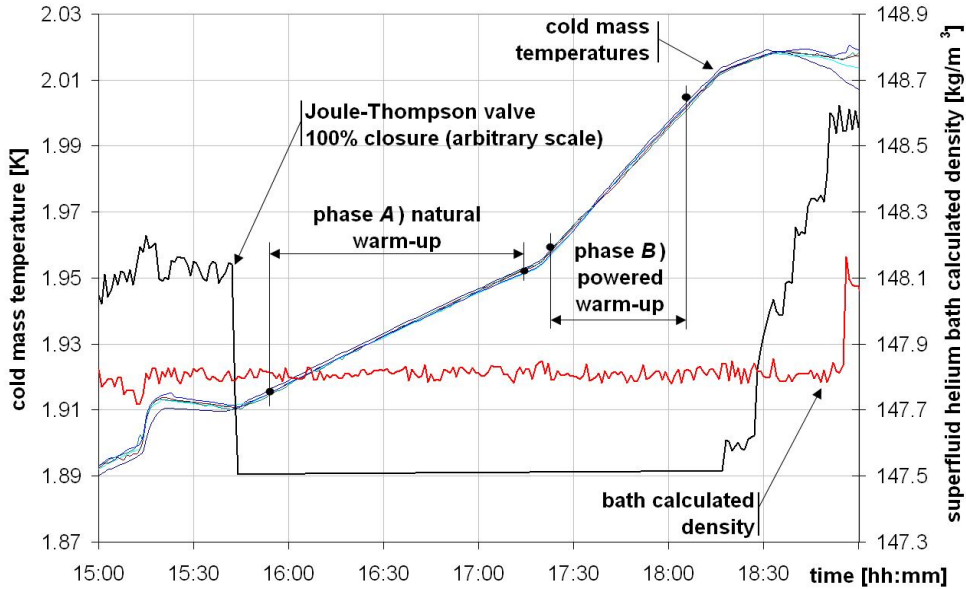

**Fig. B.3:** Temperature evolution and calculated helium density in one subsector during a test.

During this evolution, and before reaching the λ line, the set of two energy balance equations can be written:

$$
\left\{
\begin{array}{l}
W_{\mathrm{CM}} = \dfrac{\Delta U_{\mathrm{CM}}^{A}}{\Delta t^{A} \cdot L_s} = \dfrac{\sum_i m_i \cdot \Delta u_i^{A}}{\Delta t^{A} \cdot L_s} \qquad\qquad (\mathrm{B.1}) \\[4ex]
W_{\mathrm{CM}} + W_{\mathrm{IN}} = \dfrac{\Delta U_{\mathrm{CM}}^{B}}{\Delta t^{B} \cdot L_s} = \dfrac{\sum_i m_i \cdot \Delta u_i^{B}}{\Delta t^{B} \cdot L_s} \qquad\qquad (\mathrm{B.2})
\end{array}
\right.
$$

where $u = u(p,T)$ is the specific internal energy, $m_i$ and $\Delta u_i{}^j$ are the mass content and the variation of specific internal energy of the cold mass $i$-th component, respectively (different materials and unknown helium content) during the phase $j$ ($j$ being $A$ or $B$), lasting $\Delta t^j$, while $L_s$ is the subsector length.

Equation (B1) corresponds to the natural warm-up under a static load per unit length $W_{CM}$ (phase $A$), whereas Eq. (B2) corresponds to the forced warm-up, under the sum of static and electrical loads per unit length $W_{CM} + W_{IN}$ (phase $B$). Solving the system of Eqs. (1) and (2) yields the two unknowns, namely helium average content $\overline{m}_{He}$ and $W_{CM}$.

Once the helium content is determined, natural warm-ups were sufficient to measure static heat loads elsewhere in other cryogenic subsectors whenever they occurred during the LHC commissioning. Measurements were made acquiring data every minute during periods of a few hours on a number of subsectors. Insulation vacuum pressure and temperatures of lines $C'$ and $E$ were also monitored in order to assess their influence on heat loads.

Table B.1 reports test results for a number of measurement runs carried out on specific subsectors, together with parameters to which heat loads are correlated, namely temperatures of the lines $E$ and $C'$ and insulation vacuum pressure. The analysis presented hereafter assumes no sector-specific dependence of the results.

**Table B.1:** Test results for heat loads to the 1.9 K cold mass

| Test # | Test method | Cryogenic subsector/ machine sector | Cold mass heat load in subsector (mW m$^{-1}$) | Properties within subsector | | |
|--------|-------------|-------------------------------------|-------------------------------------------------|------------------|---------------------|------------------------------------|
| | | | | Line E (K) | Line C′ (K) | Insulation vacuum (Pa) |
| 1 | $A + B$ | 15_17/sector 7 right | $130.4 \pm 15.1$ | $52.2 \pm 5$ | $10 \pm 1$ | $0.9 \cdot 10^{-5}$ |
| 2 | $A + B$ | 23_25/sector 7 right | $126.5 \pm 14.7$ | $50.5 \pm 5$ | $10 \pm 1$ | $0.9 \cdot 10^{-5}$ |
| 3 | $A + B$ | 27_29/sector 7 right | $126.0 \pm 14.6$ | $49.6 \pm 5$ | $10 \pm 1$ | $4.2 \cdot 10^{-5}$ |
| 4 | $A$ | 15_13/sector 3 left | $206.8 \pm 24.0$ | $51.5 \pm 5$ | $20 \pm 1$[a] | $4.3 \cdot 10^{-5}$ |
| 5 | $A$ | 11_13/sector 2 right | $184.9 \pm 21.4$ | $43.5 \pm 5$ | $20 \pm 1$[a] | $5 \cdot 10^{-5}$ |
| 6 | $A + B$ | 17_19/sector 6 left | $133.1 \pm 15.1$ | $43.5 \pm 5$ | $7 \pm 1$ | $5.3 \cdot 10^{-5}$ |
| 7[b] | $A + B$ | 11_13/sector 5 right | $141.3 \pm 16.7$ | $53 \pm 5$ | $5.5 \pm 1$ | $3.9 \cdot 10^{-5}$ |
| 8 | $A$ | 31/sector 7 right to 33/sector 8 left | $271.6 \pm 31.5$ | $50 \pm 5$ | $6 \div 8 \pm 1$ (unstable) | $1 \cdot 10^{-4} \div 1 \cdot 10^{-3}$[c] |

[a]Line $C'$ not cooled during the test. [b]Magnet quench occurred in this subsector a few hours earlier, possibly increasing heat loads. [c]Degraded vacuum due to leaks.

From tests 1, 2, 3, 6, and 7, the average helium content has been estimated to be $26.8 \pm 3.1$ l m$^{-1}$.

Heat loads in all tests except 4, 5, 7, and 8 (discussed below) are more than 36% lower than the budget estimate of 0.21 W m$^{-1}$. This can be in part explained by the lower than nominal temperature in line $E$, thus reducing conduction heat through the support posts and radiation heat from the thermal shield to the cold mass. However test 7, which has the closest to nominal temperatures in lines $E$ and $C'$, yields a heat load which remains 32% lower than nominal. A clear correlation between temperature of line $E$ and heat load is seen by comparing tests 5 and 4, or tests 6 and 7, where an increase of 8–10 K, while keeping all other variables fixed, translates into an increase in heat load of 6% (12% when line $C'$ is also off-nominal).

When line $C'$ is not cooled (tests 4 and 5), an increase in heat loads up to 39% is observed (comparing tests 5 and 6). This large effect results from different additional heat loads, since line $C'$ intercepts heat from supports and also from other components. About one half of the additional heat load is imputable to the cold mass support posts, while the remaining contribution could be explained by the additional heat through other components, like power leads feed-through and beam-screens supports.

Finally test 8, in a subsector with degraded vacuum due to helium leaks, yields heat loads which are more than twice as high than those of test 2 where temperatures of the cryogenic lines are comparable.

**Heat load to the thermal shield**

In stationary conditions, the heat intercepted by the thermal shield is given by the enthalpy change during the non-isothermal heating of gaseous helium along line $E$. For each cryogenic subsector, the heat load to the thermal shield per unit length is:

$$W_E = \frac{\Delta H_s}{L_s} = \frac{(\dot{m}_E \cdot \Delta h_s)}{L_s} \tag{B.3}$$

where $\Delta h_s = \Delta h_s(p,T)$ is the specific mass enthalpy change along the $L_s$-long segment of line $E$ corresponding to the considered subsector, as a function of temperature and pressure readings, while $\dot{m}_E$ is the helium mass flow evaluated from the overall measured pressure drop $\Delta p$ along line $E$.

Measurements were made in stable conditions on a period of three consecutive days, at a sampling time of one minute, taking averages to reduce measurement uncertainties.

Only one test was carried out in one sector in stable condition with temperature of line $E$ between 45 and 55 K (5-10 K lower than nominal). The resulting heat load along the sector varies between 3.1 and 4.3 W m$^{-1}$, in average 18% lower than budgeted. Since the temperature of line $E$ was lower than nominal, these estimates are to be considered as upper limits.

In summary, even taking into account the uncertainties of the methods used to assess the heat loads in nominal conditions of the thermal shield and of the cold mass helium bath, the performances of the LHC cryostat are well within design and measured values on prototype strings of magnets. Heat loads at 1.9 K are about 25% lower than nominal when operating the cryostat thermal shield and heat intercepts at temperatures close to nominal. The heat loads on the thermal shield, estimated on one sector only, are also considerably below nominal (-18%), and would further decrease operating it at its nominal temperature.

These results also confirm that the performance of the LHC magnet cryostats has not been impaired by machine installation and in particular the interconnection assembly performed in the tunnel.